\documentclass[journal]{IEEEtran}

\usepackage{ucs}
\usepackage[utf8x]{inputenc}
\usepackage[cmex10]{amsmath}
\usepackage{cite, amsfonts, amssymb, amsthm, bm, bbm, graphicx, relsize, multirow, booktabs, tikz,subfigure,soul}
\usepackage[american]{babel}
\usepackage[T1]{fontenc}
\usepackage{algorithmic, algorithm}
\usepackage[multiple]{footmisc}
\theoremstyle{definition}

\theoremstyle{remark}

\renewcommand{\IEEEQED}{\IEEEQEDopen}

\newcommand{\ds}{\displaystyle}

\begin{document}

\title{Single-Carrier Modulation versus OFDM for Millimeter-Wave Wireless MIMO}
\author{Stefano Buzzi, \textit{Senior Member}, \textit{IEEE},  Carmen D'Andrea,
Tommaso Foggi, Alessandro Ugolini, and Giulio Colavolpe, \textit{Senior Member, IEEE}
\thanks{S. Buzzi and C. D'Andrea are with the Department of Electrical and Information Engineering, University of Cassino and Lazio Meridionale, I-03043 Cassino, Italy (\{buzzi, carmen.dandrea\}@unicas.it).}
\thanks{T. Foggi is with CNIT Research Unit, Department of Information Engineering, University of Parma, I-43100 Parma, Italy  (tommaso.foggi@nemo.unipr.it).}
\thanks{A. Ugolini and G. Colavolpe are with the Department of Engineering and Architecture, University of Parma, I-43100 Parma, Italy  (\{alessandro.ugolini, giulio\}@unipr.it).}
\thanks{This paper was partly presented at the 20th International ITG Workshop on Smart Antennas, Munich, Germany, March 2016.}
}
\maketitle

\begin{abstract}
This paper presents results on the achievable spectral efficiency  and on the energy efficiency for a wireless multiple-input-multiple-output  (MIMO) link operating at millimeter wave frequencies (mmWave) in a typical 5G scenario. Two different single-carrier modem schemes are considered, i.e., a traditional modulation scheme with linear equalization at the receiver, and a single-carrier modulation with cyclic prefix, frequency-domain equalization and  FFT-based  processing
at the receiver; these two schemes are compared with a conventional MIMO-OFDM transceiver structure. Our analysis jointly takes into account the peculiar characteristics of MIMO channels at mmWave frequencies, the use of hybrid (analog-digital) pre-coding and post-coding beamformers, the finite cardinality of the modulation structure, and the non-linear behavior of the transmitter power amplifiers.
Our results show that the best performance is achieved by single-carrier modulation with time-domain equalization, which exhibits the smallest loss due to the non-linear distortion, and whose performance can be further improved by using advanced equalization schemes. Results also confirm that performance gets severely degraded when the link length exceeds 90-100 meters and the transmit power falls below 0 dBW. 
\end{abstract}


\bibliographystyle{IEEEtran}

\IEEEpeerreviewmaketitle

\section{Introduction}
The adoption of carrier frequencies larger than 10 GHz will be one of the main new features  of fifth-generation (5G) wireless networks \cite{whatwillbe}, and, due to the availability of large and currently unused bandwidths, will be instrumental in delivering gigabit data-rates per users.  The use of mmWave frequencies for cellular communications has been thus deeply investigated in recent years \cite{6515173,mmWaverecent,RappaportGutierrezBen-DorMurdockQiaoTamir2013,coverageandcapacitymmWave, hybridprecodingmmwaves}, and several prototypes and test-beds showing the potentiality of mmWave frequencies for cellular applications are currently already available \cite{mmwavetestbed,kim2016feasibility}. 

One of the key questions about the use of mmWave frequencies and in general about 5G cellular systems is about the type of modulation that will be used at these frequencies. Indeed, while it is not even sure that 5G systems will use orthogonal frequency division multiplexing (OFDM) modulation at classical cellular frequencies \cite{BaBuCoMoRuUg14}\footnote{An agreement about the use
of filtered-OFDM at least in the Phase 1 of 5G systems seems however to have been reached for sub-6 GHz frequencies.}, 
there is still an open debate about the modulation type at mmWave frequencies, with 
reasons that push for 5G networks operating a single-carrier modulation (SCM) at mmWave frequencies \cite{mmWaverecent}. First of all, the propagation attenuation of mmWave frequencies makes them a viable technology only for small-cell, dense networks, where few users will be associated to any given base station, thus implying that 
 there is no compelling reason to exploit the efficient frequency-multiplexing features of OFDM,
and users may also be multiplexed in the time domain as efficiently as in the frequency domain.
Finally, 
 mmWave frequencies will be operated together with massive antenna arrays to overcome propagation attenuation. This makes digital beamforming unfeasible, since the energy required for digital-to-analog and analog-to-digital conversion would be huge. Thus, each user will have an own radio-frequency beamforming, which requires users to be separated in time rather than frequency. Otherwise stated, due to hardware complexity constraints, it may be difficult to have a fully digital (FD) beamformer that is sub-carrier dependent, while  it is simpler to have a unique beamformer for the whole available radio-frequency (RF) bandwidth that can change at the timeslot rate.

For efficient removal of the intersymbol interference induced by the frequency-selective nature of the channel, the use of SCM coupled with a cyclic prefix has been proposed, so that FFT-based processing might be performed at the receiver \cite{CP-SC}.
In \cite{Cudak13,Cudak2}, the null cyclic prefix single carrier (NCP-SC) scheme has been proposed for mmWave frequencies. The concept is to transmit a single-carrier signal, in which the usual cyclic prefix used by OFDM is replaced by nulls appended at the end of each transmit symbol.
Given the cited prohibitive hardware complexity of FD beamforming structures, 
several mmWave-specific MIMO architectures have been proposed, where signal processing is accomplished in a mixture of analog and digital domains (see, for instance, \cite{OverviewMIMO} and references therein). In particular, while FD beamforming requires one RF chain for each antenna, in hybrid (HY) structures a reduced number of RF chains is used, and beamforming is made partially in the digital domain and partially at RF frequencies, where only the signal phase (and not the amplitude) can be tuned prior to antenna transmission. 

This paper is concerned with the evaluation of the achievable spectral efficiency (ASE) and of the global energy efficiency (GEE) of SCM and OFDM schemes operating over MIMO links at mmWave frequencies. We consider three possible transceiver architectures: (a) SCM with linear zero-forcing (ZF)  equalization in the time domain for intersymbol interference removal and symbol-by-symbol detection;  (b) SCM with cyclic prefix and FFT-based processing and ZF equalization in the frequency domain at the receiver; and (c) plain MIMO-OFDM architecture for benchmarking purposes.
The ASE is computed by using the simulation-based technique for computing information-rates reported in \cite{ArLoVoKaZe06}; this technique, that has been already used in several other cases \cite{giulioTFpacking,BuzziADSL}, permits taking into account the finite cardinality of the modulation, and thus provides more accurate results than the ones that are usually reported in the literature and that refer to Gaussian signaling. The considered transceiver structures use HY pre-coding and post-coding beamforming structures, with a number of RF chains equal to the used multiplexing order - this is indeed the minimum possible number of RF chains and so the resulting structures are the one with the lowest complexity. 
Non-linear behavior of the transmit power amplifiers is also taken into account in our analysis.
We also provide an analysis of the system bit-error-rate (BER), under the assumption that low-density parity-check (LDPC) codes are used.

To the best of our knowledge, this is the first paper to provide a comprehensive study of a MIMO wireless link operating at mmWave frequencies taking \textit{simultaneously} into account effects such as (a) the clustered channel model for mmWave frequencies; (b) the non-linear distortion introduced by the power amplifier at the transmitter; (c) the use of HY analog/digital beamforming schemes; and (d) the finite cardinality of the modulation. Moreover, we detail the transceiver signal processing for the three considered modulation schemes explicitly taking into account the use of multiple antennas and the frequency-selectivity of the propagation channel. Finally, this is one of the first papers to provide results jointly on the ASE and on the GEE of the considered modulation schemes.
 
Our results will show that, among the three considered transceiver schemes, SCM with time-domain equalization (SCM-TDE) achieves  the best overall performance, since, although being slightly inferior to OFDM for the case of an ideal transmit power amplifier, it reveals to be the best option when transmitter non-linearities are taken into account. In particular, OFDM shows to be very sensitive to non-linear distortion and incurs a heavy performance degradation. We also quantify the superiority of HY beamforming with respect to FD beamforming in terms of GEE. 
Moreover, our results provide a further confirmation of the fact that   
for distances up to 100 meters, and with a transmit power around 0 dBW, mmWave links exhibit a very good performance and may be very useful in wireless cellular applications; for larger distances instead, either larger values of the transmit power or 
a larger number of antennas must be employed to overcome the distance-dependent increased attenuation.  

The rest of this paper is organized as follows. 
Next section contains the system model, with details on the considered mmWave channel model and on the front-end transmitter and receiver. In Section III the three considered transceiver structures, namely SCM-TDE, SCM with frequency-domain equalization (SCM-FDE) and MIMO-OFDM, are accurately described, while the design of the HY pre-coding and post-coding beamforming structures is reported in Section IV. Section V deals with the procedure used to compute the system ASE taking into account the finite modulation cardinality; it also models the power consumption of the considered trasceivers and provides the GEE definition.
Extensive numerical results on the system ASE, on the system GEE,  and on the coded BER are illustrated and discussed in Section VI, while,  finally, Section VII contains concluding remarks.

\medskip
\noindent\emph{Notation: }
The symbol $(\cdot)^H$ denotes conjugate transpose, $(\cdot)^T$ denotes transpose, and $\mathbf{I}_N$ denotes the $(N \times N)$-dimensional identity matrix.
The symbol $\circledast$ denotes circular convolution,  while, finally, $\| \cdot \|_{\rm F}$ denotes the Frobenius norm.

\begin{figure*}[t]
\centering
\includegraphics[scale=0.32]{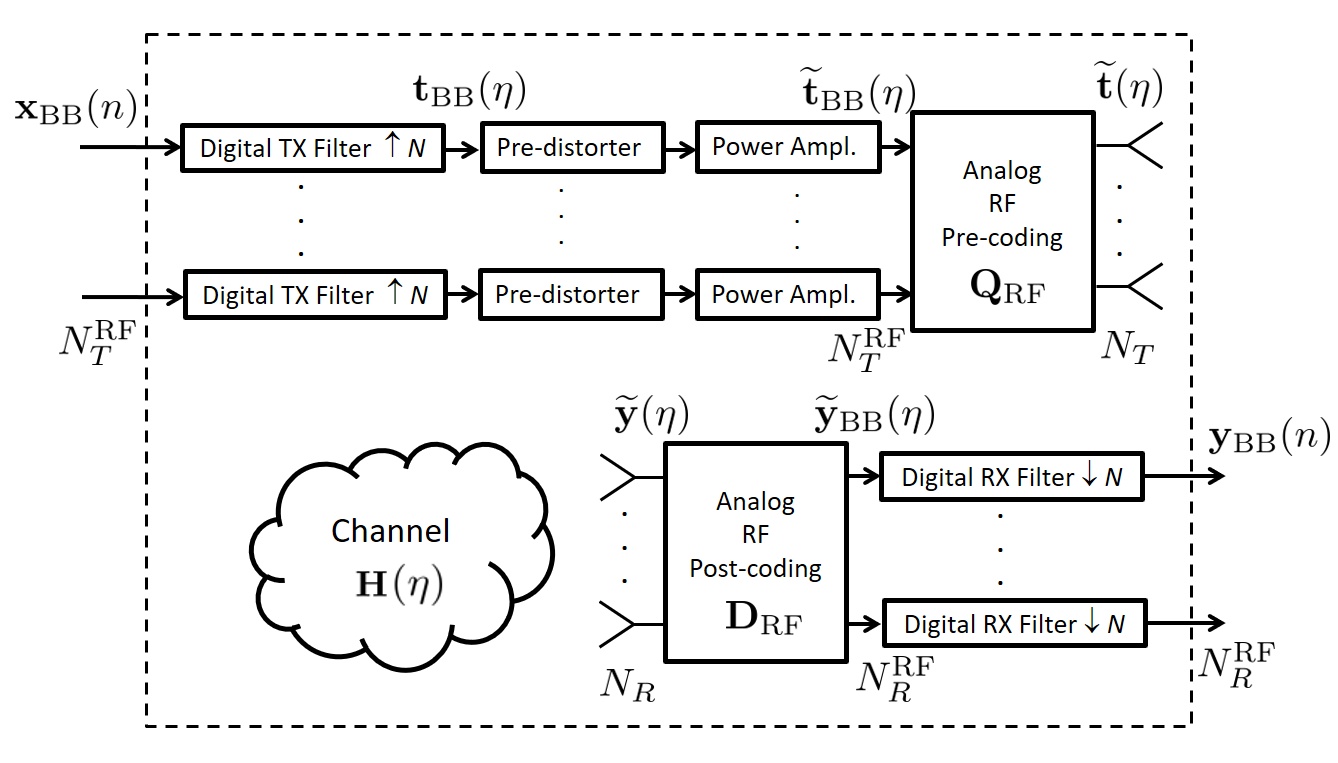}
\caption{The considered RF transceiver. It can be deemed as a non-linear system with $N_T^{\rm RF}$ inputs and $N_R^{\rm RF}$ outputs. The transceiver model also includes an SPD to compensate for the power amplifiers' non-linearities. The parameter $N$ denotes the oversampling factor used for the discrete-time approximation of the continuous-time filtering operations.}
\label{fig:scenario0}
\end{figure*}

\begin{figure*}[t]
\centering
\includegraphics[scale=0.32]{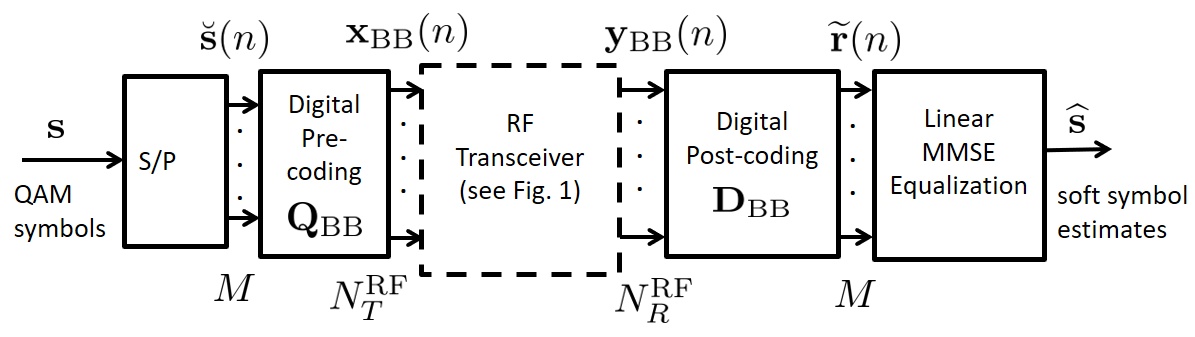}
\caption{Transceiver block-scheme for SCM with TDE.}
\label{fig:scenario1}
\end{figure*}

\section{System model}
We consider a  single-user transmitter-receiver pair that, 
 for an idealized scenario with a strictly orthogonal access scheme and no
out-of-cell interference
 may be also representative of either the uplink or the downlink of a cellular system. We denote by $N_T$ and $N_R$ the number of transmit and receive antennas, respectively, and consider the general case of a frequency-selective channel.

\subsection{The channel model}
The propagation channel can be modeled as an  $(N_R \times N_T)$-dimensional matrix-valued continuous time function, that we denote by $\mathbf{H}(t)$. 
According to the popular clustered model for MIMO mmWave channels, we assume that the propagation environment is made of $N_{\rm cl}$ scattering clusters, each of which contributes with $N_{{\rm ray}, i}$ propagation paths  $i=1, \ldots, {N_{\rm cl}}$, plus a 
possibly present LOS component. 
We denote by  $\phi_{i,l}^r$ and $\phi_{i,l}^t$ the azimuth angles of arrival and departure of the $l^{th}$ ray in the $i^{th}$ scattering cluster, respectively; similarly, $\theta_{i,l}^r$ and $\theta_{i,l}^t$ are the elevation angles  of arrival and departure of the $l^{th}$ ray in the $i^{th}$ scattering cluster, respectively. 
The impulse response of the time-continuous propagation channel  is a matrix-valued (of dimension $N_R \times N_T$) function written as
\begin{multline}
\widetilde{\mathbf{H}}(t)=\gamma\sum_{i=1}^{N_{\rm cl}}\sum_{l=1}^{N_{{\rm ray},i}}\alpha_{i,l}
\sqrt{L(r_{i,l})} \mathbf{a}_r(\phi_{i,l}^r,\theta_{i,l}^r) 
\cdot \\ 
\mathbf{a}_t^H(\phi_{i,l}^t,\theta_{i,l}^t)  \delta(t-\tau_{i,l}) + 
\widetilde{\mathbf{H}}_{\rm LOS}(t)\; .
\label{eq:channel1}
\end{multline}
In the above equation,  $\delta(\cdot)$ is the Dirac's delta,  $\alpha_{i,l}$ and $L(r_{i,l})$ are the complex path gain and the attenuation associated  to the $(i,l)$-th propagation path (whose length is denoted by $r_{i,l}$),  respectively; $\tau_{i,l}=r_{i,l}/c$, with $c$ the speed of light, is the propagation delay associated with the $(i,l)$-th path. 
The complex gain  $\alpha_{i,l}\thicksim \mathcal{CN}(0, \sigma_{\alpha,i}^2)$, with  $\sigma_{\alpha,i}^2=1$\cite{spatiallysparse_heath}. The factors $\mathbf{a}_r(\phi_{i,l}^r,\theta_{i,l}^r)$ and $\mathbf{a}_t(\phi_{i,l}^t,\theta_{i,l}^t)$ represent the normalized receive and transmit array response vectors evaluated at the corresponding angles of arrival and departure; additionally, $\gamma=\displaystyle\sqrt{\frac{N_R N_T}{\sum_{i=1}^{N_{\rm cl}}N_{{\rm ray},i}}}$  is a normalization factor ensuring that the received signal power scales linearly with the product $N_R N_T$. 
Regarding the array response vectors $ \mathbf{a}_r(\phi_{i,l}^r,\theta_{i,l}^r)$ and $\mathbf{a}_t(\phi_{i,l}^t,\theta_{i,l}^t)$, a planar antenna array configuration is used for the transmitter and receiver, with $Y_r$, $Z_r$ and $Y_t$, $Z_t$ antennas respectively on the horizontal and vertical axes for the receiver and for the transmitter. Letting $k=2\pi/\lambda$, $\lambda$ the considered wavelength, and denoting by $\tilde{d}$ the inter-element spacing we have 
$$
\begin{array}{lll}
\mathbf{a}_x(\phi_{i,l}^x,\theta_{i,l}^x)=\!\frac{1}{\sqrt{Y_xZ_x}}[1,\ldots,e^{-jk\tilde{d}(m\sin{\phi_{i,l}^x}\sin{\theta_{i,l}^x}+n\cos{\theta_{i,l}^x})},  \\
\ldots,e^{-jk\tilde{d}((Y_x-1)\sin{\phi_{i,l}^x}\sin{\theta_{i,l}^x}+(Z_x-1)\cos{\theta_{i,l}^x})}] \; ,
\end{array}
$$
where $x$ may be either $r$ or $t$.
Let us now comment on the LOS component $\widetilde{\mathbf{H}}_{\rm LOS}(t)$ in \eqref{eq:channel1}. Denoting by 
$\phi_{\rm LOS}^r$,  $\phi_{\rm LOS}^t$,
$\theta_{\rm LOS}^r$,  and $\theta_{\rm LOS}^t$ the departure angles corresponding to the LOS link, we assume that
\begin{equation}
\begin{array}{llll}
\widetilde{\mathbf{H}}_{\rm LOS}(t) = &  
I_{\rm LOS}(d) \sqrt{N_R N_T} e^{j \eta} \sqrt{L(d)}\mathbf{a}_r(\phi_{\rm LOS}^r,\theta_{\rm LOS}^r) 
\cdot \\ & 
\mathbf{a}_t^H(\phi_{\rm LOS}^t,\theta_{\rm LOS}^t) \delta(t - \tau_{\rm LOS}) \; .
\end{array}
\label{eq:Hlos}
\end{equation}
In the above equation, $\eta \thicksim \mathcal{U}(0 ,2 \pi)$, while $I_{\rm LOS}(d) $ is an {indicator function/Bernoulli random variable, equal to 1} if a LOS link exists between transmitter and receiver. We refer the reader to  \cite{buzzidandreachannel_model} for a complete specification of all the channel parameters needed to describe the channel model in \eqref{eq:channel1}. 

A comment is now in order about the frequency-selectivity of this channel. As already discussed, the strong path-loss and atmospheric absorption makes mmWave channels useful for short-range communications (up to 100-200 meters). Since the considered model relies on the direct path and on single-reflections paths, it is reasonable to assume that the differences between the lengths of the propagation paths are not larger than $10\%$ of the distance between the transmitted and the receiver.\footnote{We are actually making a conservative choice here.} Accordingly, for a link length of 150 meters, this difference is 15 meters, which results in a multipath delay spread $\tau_s$ equal to 0.05 $\mu$s. Assuming a communication bandwidth of $W=500$ MHz, the use of raised-cosine pulses with roll-off factor $\alpha=0.22$, we have a symbol-time $T_s= (1+\alpha)/W=2.44$ ns, which leads to a discrete-time channel with $\tau_s/T_s \approx 21$ taps. This number can become even larger in the case of larger communication bandwidth or in rich scattering environments. The above reasoning shows that at mmWave the MIMO wireless channel produces severe intersymbol interference.

\subsection{The RF transceiver model}
As already discussed, this paper considers three different modem schemes for MIMO mmWave systems; all of these schemes rely on the same RF transceiver scheme, that is depicted in Fig. \ref{fig:scenario0}. In order to reduce the hardware complexity, a customary trend is to consider transceivers with a number of RF chains considerably lower than the number of transmit and receive antennas, and perform beamforming partially in the digital domain and partially in the analog (RF) domain. We thus assume that there are $N_T^{\rm RF}\leq N_T$ and $N_R^{\rm RF}\leq N_R$ RF chains at the transmitter and at the receiver, respectively.\footnote{Note that assuming $N_T^{\rm RF}= N_T$ and $N_R^{\rm RF}= N_R$ we obtain the case of FD beamforming.} The RF transceiver can be thus modeled as a MIMO system with  $N_T^{\rm RF}$-dimensonal inputs 
and $N_R^{\rm RF}$-dimensional outputs; we use a discrete-time equivalent model at the symbol-rate.  At the transmitter side we have the shaping filter, followed by the amplification stage, that we model as the cascade of a non-linear signal pre-distorter (SPD) and of a non-linear power amplifier; then an analog pre-coder follows; from a mathematical point of view the analog precoder is expressed as a $(N_T \times N_T^{\rm RF})$-dimensional matrix, $\mathbf{Q}_{\rm RF}$ say, whose entries are constrained to have constant norm. At the receiver side, the $N_R$ dimensional vector goes through the analog RF  post-coding, modeled as a $(N_R \times N_R^{\rm RF})$-dimensional matrix, $\mathbf{D}_{\rm RF}$ say,  with constant-norm entries, and then is fed to a bank of receive filters matched to the ones used in transmission; the filters outputs are then sampled at symbol time. 

\subsubsection{The transmitter non-linearities}
We now provide further details about the pre-distorter and the considered non-linear power amplifier characteristic. 
Regarding the latter, we use the popular Rapp model \cite{rapp1991effects}, namely the power amplifier is a non-linear memoryless device such that the output amplitude $A$ as a function of the input amplitude $\zeta$ (assumed with unit mean square value) is expressed as
\begin{equation}
A\left(\zeta\right)=\sqrt{P_T}\frac{\zeta}{\left(1+\zeta^{2p}\right)^{1/2p}} \; ,
\end{equation}
where $P_T$ is the amplifier saturation power and for the parameter $p$ the value $2$ is used in this paper, as in \cite{Colavolpe2012}. Concerning the phase distortion introduced by the amplifier, the so-called AM/PM amplifier characteristic, there 
is actually no generally
accepted applicable model for the AM/PM characteristic, and the manufacturers only specify the maximum slope
in degrees/dB and the input level where the phase crosses 0 degree. In this paper, again in keeping with 
\cite{Colavolpe2012}, we will assume a phase offset with a slope of 2 degrees/dB when the
input level is bigger than -1.5 dB,  and no phase distortion below that level.

Regarding the pre-distorter, it is a non-linear memoryless device whose task is to attenuate the distortion introduced by the power amplifier. Following \cite{ugolini_spd}, we use a device with the input-output relationship
$y(x)=\sum_{s=0}^S g_s x |x|^{2s},
$
which is  a memoryless Volterra series of order $2S + 1$ taking into account odd order terms only. 
The complex coefficients $\mathbf{g}=\left\lbrace g_s \right\rbrace_{s=0}^S$  can be selected to minimize the mean-square error between the ideally amplified term $\sqrt{P_T}x$   and the signal at the output of the non-linear power amplifier, as shown in \cite{ugolini_spd}. In our simulations, we took $S=2$ and $[g_0, g_1, g_2]=[0.8275+0.0601i, \,   0.6335-0.0921i, \,   0.0319-0.2234i]$.

\subsubsection{The input-output relation in the linear case}
It is also useful, for the derivations in the sequel of the paper, to analyze the transceiver in the case in which the pre-distorter is absent and the amplifier is ideal, i.e. with an AM/AM characteristic $A(\zeta)=\sqrt{P_T} \zeta$ and with no-phase distortion. In this case, it is easily recognized that the RF transceiver in Fig. \ref{fig:scenario0} reduces to a linear time-invariant LTI filter with $(N_R^{\rm RF} \times N_T^{\rm RF})$-dimensional matrix valued impulse response. 
In particular, denoting by $h_{\rm TX}(t)$ the baseband equivalent transmit shaping 
filters,\footnote{We have $N_T^{\rm RF}$ of such filters.} by 
$h_{\rm RX}(t)$ the baseband equivalent of the impulse response of the $N_R^{\rm RF}$ receive filters, and by $h(t)=h_{\rm TX}(t) \ast h_{\rm RX}(t)$ their convolution, and assuming a sampling interval equal to $T_s$, 
the RF transceiver block impulse-response of the linear time-invariant system consisting of the $N_T$ transmit shaping filters, the propagation channel, and the $N_R$ receive filters  is a matrix-valued (of dimension $N_R^{\rm RF} \times N_T^{\rm RF}$) discrete-time sequence that  can be written as follows:
\begin{equation}
{\mathbf{L}}(n)=\mathbf{D}_{\rm RF}^H \widetilde{\mathbf{H}}(n) \mathbf{Q}_{\rm RF} \; ,
\label{eq:RF transceiver}
\end{equation}
with $\widetilde{\mathbf{H}}(n) $ the $(N_R \times N_T)$-dimensional discrete-time composite channel response including the transmit and receive shaping filters: 
\begin{equation}
\begin{array}{llll}
\widetilde{\mathbf{H}}(n)=& \gamma \displaystyle \sum_{i=1}^{N_{\rm cl}}\sum_{l=1}^{N_{{\rm ray},i}}\alpha_{i,l}
\sqrt{L(r_{i,l})} \mathbf{a}_r(\phi_{i,l}^r,\theta_{i,l}^r)  \mathbf{a}_t^H(\phi_{i,l}^t,\theta_{i,l}^t) 
\cdot \\ 
&  h(nT_s-\tau_{i,l}) + 
\widetilde{\mathbf{H}}_{\rm LOS}(n)\; ,
\end{array}
\label{eq:channel_composite}
\end{equation}
with
\begin{equation}
\begin{array}{llll}
\widetilde{\mathbf{H}}_{\rm LOS}(n) = &  
I_{\rm LOS}(d) \sqrt{N_R N_T} e^{j \eta} \sqrt{L(d)}\mathbf{a}_r(\phi_{\rm LOS}^r,\theta_{\rm LOS}^r) \cdot  \\ & 
\mathbf{a}_t^H(\phi_{\rm LOS}^t,\theta_{\rm LOS}^t) h(nT_s - \tau_{\rm LOS}) \; .
\end{array}
\label{eq:Hlos_discrete}
\end{equation}
Assuming that the multipath delay spread spans $P$ sampling intervals and that the duration of the transmit and receive shaping filters spans $P_h$ sampling intervals each, it is easily seen that the matrix-valued channel sequences 
${\mathbf{L}}(n)$ and $\widetilde{\mathbf{H}}(n)$ have $\widetilde{P}=P+2P_h-1$ non-zero elements; for ease of notation, we assume, as usually happens, that the non-zero elements of ${\mathbf{L}}(n)$ and $\widetilde{\mathbf{H}}(n)$ are those corresponding to $n=0, \ldots, \widetilde{P}-1$.
Denoting by $\mathbf{x}_{\rm BB}(n)$ the $N_T^{\rm RF}$-dimensional vector at the input of the RF transceiver at discrete epoch $n$, it is easily shown that the corresponding output $\mathbf{y}_{\rm BB}(n)$ is represented by the following $N_R^{\rm RF}$-dimensional vector
\begin{equation}
\begin{array}{llllll}
\mathbf{y}_{\rm BB}(n)&=& \displaystyle \sum_{\ell=0}^{\widetilde{P}-1}\sqrt{P_T}{\mathbf{L}}(\ell) \mathbf{x}_{\rm BB}(n-\ell) + \mathbf{D}_{\rm RF}^H\mathbf{w}(n)  \\
&=&  \displaystyle \sum_{\ell=0}^{\widetilde{P}-1}\sqrt{P_T}\mathbf{D}_{\rm RF}^H\widetilde{\mathbf{H}}(\ell) \mathbf{Q}_{\rm RF}\mathbf{x}_{\rm BB}(n-\ell) + \mathbf{D}_{\rm RF}^H\mathbf{w}(n)
\, ,
\end{array}
\label{eq:io}
\end{equation}
with $\mathbf{w}(n)$ denoting the $N_R$-dimensional thermal noise vector at the output of the receive shaping filters. It is seen from \eqref{eq:io} that the input-output relationship introduces intersymbol interference (ISI), thus implying that for SCM schemes properly equalization structures will be needed. 
Regarding the additive thermal noise, it is uncorrelated across antennas, i.e., the noise samples collected through different receive antennas are statistically independent: the vector $\mathbf{w}(n)$ is thus a complex zero-mean Gaussian random variable with covariance matrix $\sigma^2_w \mathbf{I}_{N_R}$, 
with $\sigma^2_w= 2 {\cal N}_0 \int_{-\infty}^{+\infty}|h_{\rm RX}(t)|^2 {\mathrm{d}t}$. Conversely, noise samples are in general correlated through time, i.e., we have
$E\left[ w_i(n) w_i^*(n-l)\right]=2 {\cal N}_0  r_{ h_{\rm RX}}(lT_s) \; , 
$ $\forall i=1, \ldots, N_R$, where $w_i(n)$ denotes the $i$-th entry of the vector $\mathbf{w}(n)$, and  $r_{ h_{\rm RX}}(\tau)=\int_{-\infty}^{+\infty} h_{\rm RX}(t) h^*_{\rm RX}(t-\tau){\mathrm{d}t}$ denotes the correlation function of the receive shaping filter. It thus follows that, if we arrange $L$ consecutive noise vectors in an $(N_R \times L)$-dimensional matrix
$
\mathbf{W}=[ \mathbf{w}(n) \; \mathbf{w}(n-1) \; \ldots, \mathbf{w}(n-L+1)]\; , 
$ 
we have that the entries of the matrix $\mathbf{W}$ are vertically uncorrelated (actually, independent) and horizontally correlated.


\section{Transceiver processing}
In the following, we illustrate the signal processing operations performed by the three considered transceiver schemes, namely SCM-TDE, SCM-FDE, and OFDM. We have already commented on the RF transceiver block of Fig. \ref{fig:scenario0}, so, in the following, the emphasis will be on the description of the remaining blocks. Since, for the sake of simplicity, we will not consider any processing at the receiver aimed at taking care of the transmitter non-linearities,\footnote{The non-linearity of the power amplifier is taken into account at the transmitter through the pre-distortion and a possible input backoff.} we will describe the signal processing at the receiver under the assumption that the whole system is linear. Of course, when performing numerical simulation, we will include, where needed, the effect of non-linearities. 
This will permit assessing the effect that non-linear power amplifiers have on the system performance when the receiver has been designed under the assumption of perfectly linear transmitter.
The design of receiver schemes explicitly taking into account the power amplifier non-idealities is indeed a research topic left for future work.

Denote now by $\mathbf{s}$ a column vector containing the $L$ data-symbols -- drawn either from a QAM constellation or from a Gaussian distribution, and with unit average energy  -- to be transmitted:
$\mathbf{s}=[s_0, s_1, \ldots, s_{L-1}]^T \; .
$
We assume that $L=kM$, where $k$ is an integer and  $M$, the multiplexing order, is the number of information symbols that are 
simultaneously transmitted by the $N_T$ transmit antennas in each symbol interval. 
In the following, we present three possible transceiver models. 

\subsection{SCM with TDE}

We refer to the discrete-time block-scheme reported in Fig.~\ref{fig:scenario1}. 
The QAM symbols in vector $\mathbf{s}$ are fed to a serial-to-parallel {(S/P)} conversion block that splits them in $k$ distinct $M$-dimensional vectors $\breve{\mathbf{s}}(1) , \ldots, \breve{\mathbf{s}}(k) $. These vectors are pre-coded using 
the $(N_T^{\rm RF} \times M)$-dimensional digital pre-coding matrix $\mathbf{Q}_{\rm BB}$; 
we thus obtain the $N_T^{\rm RF}$-dimensional vectors 
$
\mathbf{x}_{\rm BB}(n)=\mathbf{Q}_{\rm BB}\breve{\mathbf{s}}(n)$, $n=1, \ldots, k $.
The vectors $\mathbf{x}_{\rm BB}(n)$ are fed to a bank of $N_T^{\rm RF}$ identical shaping filters, converted to RF, amplified, pre-coded at RF and transmitted. 

At the receiver, after RF post-coding and baseband-conversion, the $N_R^{\rm RF}$ received signals are passed through a bank of filters matched to those used for transmission and sampled at symbol-rate. We thus obtain the $N_R^{\rm RF}$-dimensional vectors $\mathbf{y}_{\rm BB}(n)$, which are passed through a digital post-coding matrix, that we denote by  $\mathbf{D}_{\rm BB}$,  of dimensions $(N_R^{\rm RF} \times M)$. 
Recalling that $\widetilde{\mathbf{H}}(n)$ is the matrix-valued finite-impulse-response (FIR) filter representing the composite channel impulse response (i.e., the convolution of the transmit filter, actual matrix-valued channel, and receive filter) it is easy to show, by virtue of the input-output relationship \eqref{eq:io} that the generic $M$-dimensional vector at the output of the post-coding matrix, say $\widetilde{\mathbf{r}}(n)$,  is written as
\begin{equation}
\begin{array}{llllll}
\widetilde{\mathbf{r}}(n) &=& \mathbf{D}_{\rm BB}^H \mathbf{y}_{\rm BB}(n) \\
&=& \displaystyle \sum_{\ell=0}^{\widetilde{P}-1}
\mathbf{D}_{\rm BB}^H {\mathbf{L}}(\ell)\mathbf{Q}_{\rm BB}\breve{\mathbf{s}}(n-\ell) + \mathbf{D}_{\rm BB}^H \mathbf{D}_{\rm RF}^H\mathbf{w}(n) \; .
\end{array}
\label{eq:received_signal_1}
\end{equation}
So far, the choice of the pre-coding and post-coding beamforming matrices $\mathbf{Q}_{\rm BB}$, $\mathbf{Q}_{\rm RF}$, $\mathbf{D}_{\rm BB}$, and $\mathbf{D}_{\rm RF}$ has been left unspecified. We now describe the considered  beamforming structures in the FD case (i.e., assuming a number of RF chains equal to the number of antennas, both at the transmitter and at the receiver, and removing the RF beamforming matrices) 
deferring to Section IV
the exposition of the algorithms for the design of the HY structures. Letting $\mu={\arg}\max_{\ell = 0, \ldots, \widetilde{P}-1}\left\{ \left\|\widetilde{\mathbf{H}}(\ell)\right\|_{\rm F}\right\}$, 
we assume here that $\mathbf{Q}_{\rm BB}$ contains on its columns the left eigenvectors of the matrix $\widetilde{\mathbf{H}}(\mu)$ corresponding to the $M$ largest eigenvalues, and that the matrix $\mathbf{D}_{\rm BB}$ contains on its columns the corresponding right eigenvectors. Note that, due to the presence of ISI, the proposed pre-coding and post-coding structures are not optimal. Nevertheless, we make here this choice for the sake of simplicity, 
 and also to avoid increasing the computational complexity gap with the OFDM scheme, resorting
 to the use of an equalizer to cancel the effects of ISI.
We will adopt a linear ZF equalizer making a block processing of $\widetilde{P}$ consecutive received data vectors: to obtain a soft estimate of the data vector $\breve{\mathbf{s}}(n)$, the $\widetilde{P}$ observables 
$\widetilde{\mathbf{r}}(n+\widetilde{P}-1) \ldots \widetilde{\mathbf{r}}(n)$
are stacked into a single $\widetilde{P}M$-dimensional vector, expressed as $
\widetilde{\mathbf{r}}_{\widetilde{P}}(n)=[\widetilde{\mathbf{r}}(n+\widetilde{P}-1) \ldots \widetilde{\mathbf{r}}(n)]^T \;.
$
Through ordinary algebra, it is easy to recognize that this vector can be expressed in the form
\begin{equation}
\widetilde{\mathbf{r}}_{\widetilde{P}}(n)=\mathbf{A}\widetilde{\mathbf{s}}_{\widetilde{P}}(n)+\mathbf{B}\widetilde{\mathbf{w}}_{\widetilde{P}}(n) \;,
\label{eq:vector_r}
\end{equation}
where $\widetilde{\mathbf{s}}_{\widetilde{P}}(n)$ is an  $M(2\widetilde{P}-1)$-dimensional vector  containing the data symbols contributing to  $\widetilde{\mathbf{r}}_{\widetilde{P}}(n)$, i.e.:
$\breve{\mathbf{s}}_{\widetilde{P}}(n)=[\breve{\mathbf{s}}(n+\widetilde{P}-1) \ldots \breve{\mathbf{s}}(n)\ldots \breve{\mathbf{s}}(n-\widetilde{P}+1)]^T 
$,
$\widetilde{\mathbf{w}}_{\widetilde{P}}(n)=[\widetilde{\mathbf{w}}(n+\widetilde{P}-1) \ldots \widetilde{\mathbf{w}}(n)]^T 
$
 is the $N_R\widetilde{P}$-dimensional noise vector
and  $\mathbf{A}$ and $\mathbf{B}$ are suitable matrices, of dimension  $[M\widetilde{P}\times M(2\widetilde{P}-1)]$ and  $[M\widetilde{P}\times N_R\widetilde{P}]$, respectively. 
The ZF-equalized soft estimate  of the desired data vector $\widehat{\breve{\mathbf{s}}}(n)$ is obtained through the following processing:
\begin{equation}
\widehat{\breve{\mathbf{s}}}(n)=\mathbf{E}^H\widetilde{\mathbf{r}}_{\widetilde{P}}(n) \; ,
\label{eq:processingTDE}
\end{equation}
where  $\mathbf{E}$ is the $(\widetilde{P}M \times M)$-dimensional ZF equalizer. Its $\ell$-th column  is expressed as \cite{kay1998}:
\begin{equation}
\mathbf{E}(:,\ell)=\ds \frac{(\mathbf{A}\mathbf{A}^H)^+ \mathbf{A}(:,M(\widetilde{P}-1)+\ell)}{\|(\mathbf{A}\mathbf{A}^H)^+ \mathbf{A}(:,M(\widetilde{P}-1)+\ell)\|_F}\; ,
\end{equation}
with $\ell=1, \ldots, M$, and $(\cdot)^+$ denoting Moore-Penrose pseudo-inverse.
Note that we are here considering a ZF equalizer based on the use of the pseudo-inverse: this equalizer performs similarly to the MMSE equalizer in the case in which the interference spans the whole signal space and no interference-free signal space dimensions exist. Another possible approach is instead to perform signal projection onto the interference-free signal space; in the current scenario, however, using this latter approach would require the consideration of  a fractionally-spaced equalizer in order to have a non-empty interference-free subspace. For the sake of simplicity, we thus choose to adopt the former approach.

\noindent
{\em Considerations on complexity.} Regarding processing complexity, we note that the computation of the equalization matrix $\mathbf{E}$ requires  a computational burden proportional to $(\widetilde{P}M)^3$;
then,  implementing \eqref{eq:processingTDE} requires a matrix vector product, with a computational burden proportional to $(\widetilde{P} M^2)$; this latter task must be made $k$ times in order to provide the soft vector estimates for all values of $n=1, \ldots, k.$

\begin{figure*}[!t]
\centering
\includegraphics[scale=0.32]{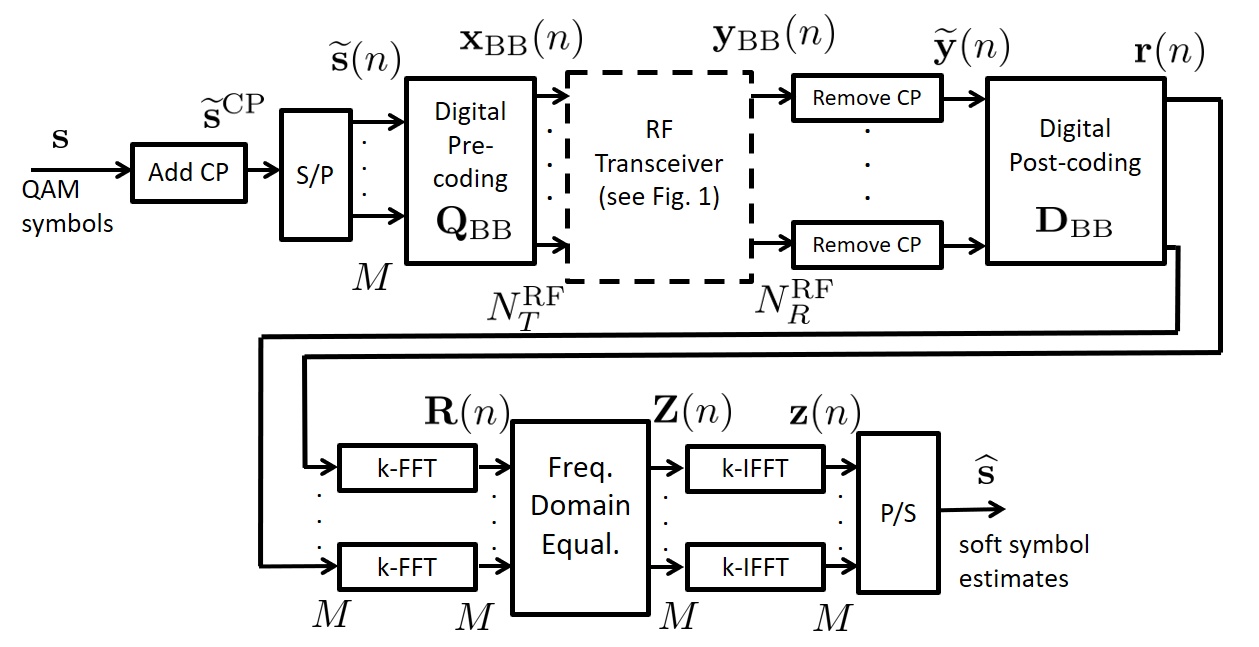}
\caption{Transceiver block-scheme for SCM with cyclic prefix, FFT-based processing and FDE.}
\label{fig:scenario2}
\end{figure*}

\subsection{SCM with FDE}
We now consider the case in which SCM is used in conjunction with a CP and FDE; 
we refer to the discrete-time block-scheme reported in Fig. \ref{fig:scenario2}. 
A cyclic prefix of length $CM$ is added at the beginning of the block $\mathbf{s}$ of $L=kM$ QAM symbols, so as to have the vector $\widetilde{\mathbf{s}}$ of {length} $(k+C)M$.
As in the previous case, the vector $\widetilde{\mathbf{s}}$ is passed through a serial-to-parallel conversion with $M$ outputs, a digital pre-coding block (again expressed through the matrix $\mathbf{Q}_{\rm BB}$), a bank of $N_T^{\rm RF}$ transmit filters; then conversion to RF, signal amplification, analog RF pre-coding and transmission take place.
At the receiver, after RF post-coding and baseband-conversion, the $N_R^{\rm RF}$ received signals are passed through a bank of filters matched to the ones used for transmission and sampled at symbol-rate; then, the cyclic prefix is removed.  We thus obtain the $N_R^{\rm RF}$-dimensional vectors
$\widetilde{\mathbf{y}}(n)$, with $n=1, \ldots, k$, containing a noisy version of the \textit{circular} convolution between the sequence $\widetilde{\mathbf{x}}(n) \triangleq 
\mathbf{Q}_{\rm RF} \mathbf{x}_{\rm BB}(n)$ and $\widetilde{\mathbf{H}}(n)$ , i.e.:
\begin{equation}
\widetilde{\mathbf{y}}(n)=
\sqrt{P_T} \mathbf{D}^H_{\rm RF} \left[\widetilde{\mathbf{H}}(n) \circledast \widetilde{\mathbf{x}}(n)\right]  + \mathbf{w}(n) \; , \qquad n=1, \ldots, k
\end{equation}
The vectors $\widetilde{\mathbf{y}}(n)$ are then 
processed by the digital post-coding matrix $\mathbf{D}_{\rm BB}$. The choice of the beamforming matrices  is the same as that of the previous subsection (SCM with TDE) for the FD case, so we do not comment on it here; again, the case of HY analog/baseband beamforming is treated in the sequel of the paper. 
After digital post-coding beamforming, we obtain the $M$-dimensional vectors $\mathbf{r}(n)=\mathbf{D}_{\rm BB}^H \widetilde{\mathbf{y}}(n)$, with $n=1, \ldots, k$.
These vectors go through an entry-wise FFT transformation on $k$ points; the $n$-th FFT coefficient, with $n=1, \ldots, k$, can be shown to be expressed as\footnote{The factor $\sqrt{k}$ in \eqref{eq:fftdomain} is due to the fact that we are considering isometric FFT and IFFT transformations.}
\begin{equation}
\mathbf{R}(n)= \sqrt{k P_T} \widetilde{{\cal H}}(n) \mathbf{X}(n) + \mathbf{W}(n) \; ,
\label{eq:fftdomain}
\end{equation}
where $\widetilde{{\cal H}}(n)$ is an $(M \times N_T)$-dimensional matrix representing the $n$-th FFT coefficient of the matrix-valued sequence $\mathbf{D}_{\rm BB}^H \mathbf{D}_{\rm RF}^H \widetilde{\mathbf{H}}(n)$, and $ \mathbf{X}(n)$ and $\mathbf{W}(n)$ are the $n$-th FFT coefficient of the sequences $\widetilde{\mathbf{x}}(n)$ and $\mathbf{D}_{\rm BB}^H \mathbf{D}_{\rm RF}^H \mathbf{w}(n)$, respectively.
From (\ref{eq:fftdomain}), it is seen that, due to the presence of multiple antennas, and, thus, of the matrix-valued channel, the useful symbols reciprocally interfere and an equalizer is needed. \eqref{eq:fftdomain} can be also shown to be expressed as:
\begin{equation}
\mathbf{R}(n)= \sqrt{k P_T} \widetilde{{\cal H}}(n) 
\mathbf{Q}_{\rm RF} \mathbf{Q}_{\rm BB}
\widetilde{\mathbf{S}}(n) + \mathbf{W}(n) \; ,
\label{eq:fftdomain2}
\end{equation}
with $\widetilde{\mathbf{S}}(n)$ an $M$-dimensional vector representing the $n$-th FFT coefficient of the vector-valued sequence $\widetilde{\mathbf{s}}(n)$.\footnote{We used here the relation 
$\mathbf{X}(n)= \mathbf{Q}_{\rm RF} \mathbf{Q}_{\rm BB}
 \widetilde{\mathbf{S}}(n)$.}
We denote by $\mathbf{E}(n)$ the $(M \times M)$-dimensional equalization matrix, and a  zero-forcing approach is adopted, thus implying that  $\mathbf{E}^H(n) =  (\sqrt{k P_T}\widetilde{{\cal H}}(n) \mathbf{Q}_{\rm RF} \mathbf{Q}_{\rm BB})^{-1}$. The output of the equalizer is written as
$$
\mathbf{Z}(n)= \mathbf{E}^H(n) \mathbf{R}(n)=\widetilde{\mathbf{S}}(n) +  
 (\sqrt{k P_T}\widetilde{{\cal H}}(n) \mathbf{Q}_{\rm RF} \mathbf{Q}_{\rm BB})^{-1}\mathbf{W}(n) \; .
$$
Then, the vectors $\mathbf{Z}(n)$ go through an entry-wise IFFT transformation on $k$ points. It can be shown that the $n$-th IFFT coefficient of the vector $\mathbf{Z}(n)$ can be expressed as:
\begin{equation}
\mathbf{z}(n)=\widetilde{\mathbf{s}}(n)+ \left[\mathbf{I}_{M} \otimes \left[\mathbf{D}_{\rm IFFT}\right]_{:,n}\right] \mathbf{N}_{ \rm stacked} \;,
\end{equation}
where  $\left[\mathbf{D}_{\rm IFFT}\right]_{:,n}$ is the $n$-th column of the isometric IFFT matrix  $\mathbf{D}_{\rm IFFT}$, whose $(m,l)$-th element is given by 
$
\mathbf{D}_{\rm IFFT}{(m,l)}=\displaystyle \frac{1}{\sqrt{k}}e^{j 2\pi\frac{(m-1)(l-1)}{k}} $,
and $\mathbf{N}_{ \rm stacked}$ is the $kM$-dimensional vector  containing the stacked vectors 
 $(\sqrt{k P_T}\widetilde{{\cal H}}(1) \mathbf{Q}_{\rm RF} \mathbf{Q}_{\rm BB})^{-1}\mathbf{W}(1),$ $\ldots,$ $(\sqrt{k P_T}\widetilde{{\cal H}}(k) \mathbf{Q}_{\rm RF} \mathbf{Q}_{\rm BB})^{-1}\mathbf{W}(k)$.

\noindent
{\em Considerations on complexity.} Looking at the scheme in Fig. \ref{fig:scenario2}, the computational burden of the considered transceiver architecture is the following. $2M$ FFTs of length $k$ are required, with a complexity proportional to $2M k \log_2 k$; in order to compute the zero-forcing matrix, the FFT of the matrix-valued sequence $\widetilde{{\cal H}}(n)$ must be computed, with a complexity proportional to $MN_T (k \log_2 k)$; computation of the matrix $(\widetilde{{\cal H}}(n) \mathbf{Q}_{\rm RF} \mathbf{Q}_{\rm BB})$ and of its inverse, for $n=1, \ldots, k$,  finally requires a computational burden proportional to $k(N_T M^2 + M^3)$. 
It can be easily seen that the complexity of the FDE scheme is lower than that of the TDE scheme.

\begin{figure*}[!t]
\centering
\includegraphics[scale=0.32]{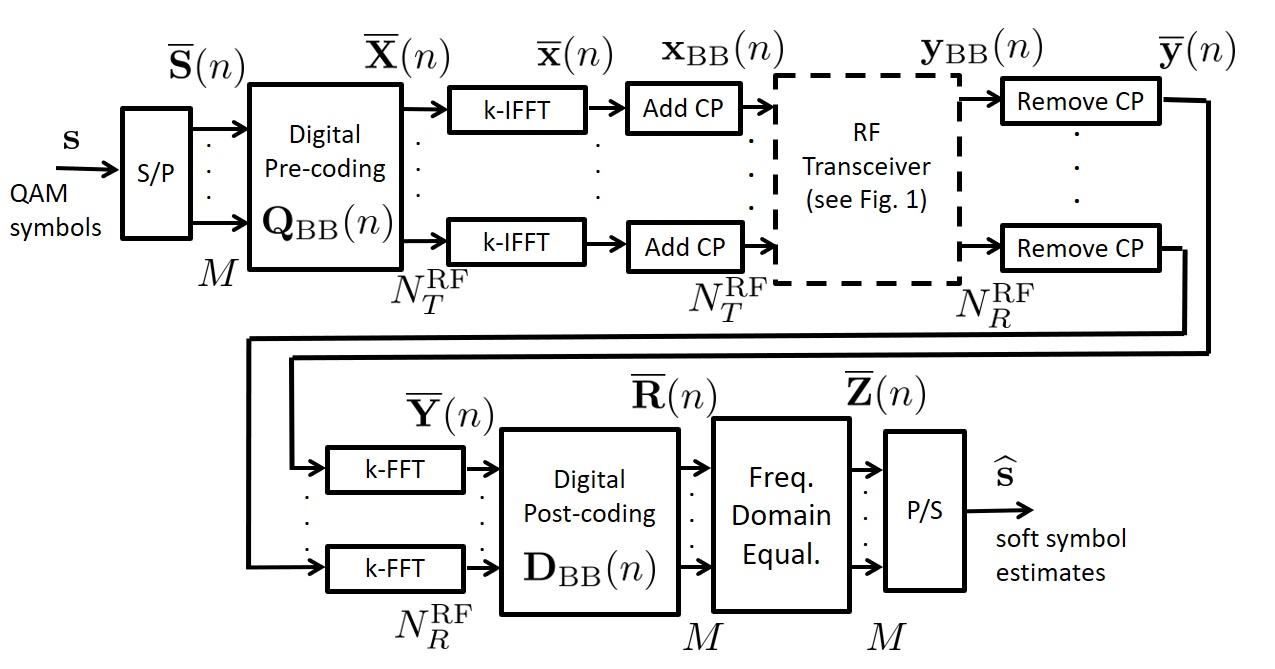}
\caption{Transceiver block-scheme for OFDM with FDE.}
\label{fig:OFDMscheme}
\end{figure*}

\subsection{Transceiver model - OFDM}
Finally, we consider the 
MIMO-OFDM discrete-time block-scheme reported in Fig. \ref{fig:OFDMscheme}. Differently from previous schemes, baseband beamforming is made on a "per-subcarrier" basis  \cite{castanedaWSA2016}, while the analog beamformer jointly process the entire signal bandwidth, i.e., all the subcarriers are treated uniformly. 

Each OFDM symbol is assumed to be  made of $L=kM$ QAM data symbols; after S/P conversion, the data symbols are split in $k$ distinct $M$-dimensional vectors $\overline{\mathbf{S}}(1) , \ldots, \overline{\mathbf{S}}(k) $. These vectors are pre-coded through the $(N_T^{\rm RF} \times M)$-dimensional digital pre-coding matrices $\mathbf{Q}_{\rm BB}(1), \ldots , \mathbf{Q}_{\rm BB}(k)$,  thus yielding the vectors $\overline{\mathbf{X}}(n)= \mathbf{Q}_{\rm BB}(n)\overline{\mathbf{S}}(n)$. These vectors then go through an entry-wise IFFT transformation on $k$ points; we denote by $\overline{\mathbf{x}}(n)$ the $M$-dimensional transformed vectors, with $n=1, \ldots, k$. 
A CP of length $C$ is added at the beginning of the block so that we have the following sequence of $N_T^{\rm RF}$-dimensional vectors:
\begin{equation}
\mathbf{x}_{\rm BB}(n)=\left\{ \begin{array}{llll}
\overline{\mathbf{x}}(n+k-C) \; , & n=1, \ldots, C \; , \\
\overline{\mathbf{x}}(n-C) \; , & n=C+1, \ldots, C+k \; .
\end{array}\right.
\end{equation}
The vectors $\mathbf{x}_{\rm BB}(n)$ are passed through a bank of $N_T^{\rm RF}$ transmit filters, converted to RF,
amplified,
RF-precoded, and transmitted.
At the receiver, analog RF post-coding, baseband-conversion, matched-filtering and sampling at the symbol-rate, 
the cyclic prefix is removed.  We thus obtain the following $N_R^{\rm RF}$-dimensional vectors
$\overline{\mathbf{y}}(n)$, with $n=1, \ldots, k$:
\begin{equation}
\overline{\mathbf{y}}(n)=\sqrt{P_T} \mathbf{D}_{\rm RF}^H \left[
 \widetilde{\mathbf{H}}(n) \circledast \mathbf{Q}_{\rm RF}\overline{\mathbf{x}}(n)\right]  + \mathbf{D}_{\rm RF}^H \mathbf{w}(n) \; , 
\end{equation}
with $\widetilde{\mathbf{H}}(n)$  denoting again the matrix-valued FIR filter representing the composite channel impulse response
(i.e., the convolution of the transmit filter, actual matrix-valued channel and receive filter).
These vectors go through an entry-wise FFT transformation on $k$ points; the $n$-th FFT coefficient, with $n=1, \ldots, k$, can be shown to be expressed as
\begin{equation}
\overline{\mathbf{Y}}(n)= \sqrt{k P_T}\mathbf{D}_{\rm RF}^H \overline{{\cal H}}(n)\mathbf{Q}_{\rm RF}\mathbf{Q}_{\rm BB}(n)\overline{\mathbf{S}}(n)  + \mathbf{D}_{\rm RF}^H \overline{\mathbf{W}}(n) \; ,
\label{eq:fftdomain_OFDM}
\end{equation}
where $\overline{{\cal H}}(n)$ is an $(N_R \times N_T)$-dimensional matrix representing the $n$-th FFT coefficient of the matrix-valued sequence $\widetilde{\mathbf{H}}(n)$, and $\overline{\mathbf{W}}(n)$ is the $n$-th FFT coefficient of the sequence $\mathbf{w}(n)$, respectively.
The vectors $\overline{\mathbf{Y}}(n)$ 
are then processed by the digital post-coding matrix $\mathbf{D}_{\rm BB}(n)$; we thus obtain the $M$-dimensional vectors 
\begin{equation}
\begin{array}{lll}
\overline{\mathbf{R}}(n)=& \sqrt{k P_T}\mathbf{D}_{\rm BB}^H(n) 
\mathbf{D}_{\rm RF}^H \overline{{\cal H}}(n)\mathbf{Q}_{\rm RF}\mathbf{Q}_{\rm BB}(n)\overline{\mathbf{S}}(n)  +  \\ & 
\mathbf{D}_{\rm BB}^H(n) \mathbf{D}_{\rm RF}^H \overline{\mathbf{W}}(n) \; , \; \;  n=1, \ldots, k \; .
\end{array}
\label{eq:fftdomain_OFDM22}
\end{equation}
From \eqref{eq:fftdomain_OFDM}, it is seen that, due to the presence of multiple antennas, and, thus, of the matrix-valued channel, the useful symbols reciprocally interfere and thus an equalizer is needed.
Denoting by $\overline{\mathbf{E}}(n)$ the $(M \times M)$-dimensional equalization matrix, and using a zero-forcing approach, it can be seen that  $\overline{\mathbf{E}}^H(n) =  (\sqrt{k P_T} \mathbf{D}_{\rm BB}^H(n) \mathbf{D}_{\rm RF} \overline{{\cal H}}(n) \mathbf{Q}_{\rm RF} \mathbf{Q}_{\rm BB}(n))^{+}$, where $(\cdot)^+$ denotes the Moore-Penrose pseudoinverse. The output of the equalizer can be shown to be {expressed} as:
\begin{multline}
\overline{\mathbf{Z}}(n)= \overline{\mathbf{E}}(n)^H \overline{\mathbf{R}}(n)=\overline{\mathbf{S}}(n) + \\    
(\sqrt{k P_T}\mathbf{D}_{\rm BB}(n)^H \mathbf{D}_{\rm RF}^H \overline{{\cal H}}(n) \mathbf{Q}_{\rm RF} \mathbf{Q}_{\rm BB}(n))^{+} \mathbf{D}_{\rm BB}(n)^H \overline{\mathbf{W}}(n) \; .
\end{multline}
After P/S conversion we finally obtain the soft estimates of the transmitted symbols.

\section{HY architecture design}

%

At mmWave antenna arrays usually have more elements than at lower frequencies, due also to the small size induced by the reduced wavelength. As already said, current technology prevents the use of FD beamformers and thus signal processing operations are shared among the digital and analog domains; lower-resolution 
data-converters are also used \cite{OverviewMIMO}. HY architectures are one approach for providing enhanced benefits of MIMO communication at mmWave frequencies \cite{spatiallysparse_heath}.  Similar solutions are also reported in \cite{alkhateeb2014channel,kim2013tens}.

A small number of transceivers is assumed, so that $M < N_T^{RF} < N_T$ and $M < N_R^{RF} < N_R$. 
We now detail the low-complexity HY beamforming structures. We have already specified the beamforming matrices for the FD case. Now, these beamforming matrices are to be approximated through the cascade of an analog and baseband precoder. In the following, we denote by $\mathbf{Q}^{\rm opt}$ and $\mathbf{D}^{\rm opt}$ the beamforming pre-coding and post-coding structures (specified in the previous section for all the considered modulation schemes) to be approximated through the HY structure.
We first deal with the case of SCM, and then we will examine the MIMO-OFDM case.

\subsection{Hybrid beamforming for SCM schemes}

In order to reduce hardware complexity with respect to the FD beamforming, in HY structures the 
$(N_T \times M)-$dimensional
pre-coding matrix  is written as the product $\mathbf{Q}_{ \rm RF}\mathbf{Q}_{\rm BB}$, where $\mathbf{Q}_{ \rm RF}$ is the $(N_T \times N_T^{RF})$-dimensional RF pre-coding matrix and $\mathbf{Q}_{\rm BB}$ is the $(N_T^{RF} \times M)-$dimensional baseband pre-coding matrix. Since the RF precoder is implemented using phase shifters, the entries of the matrix $\mathbf{Q}_{RF}$ 
have all the same magnitude (equal to $\frac{1}{\sqrt{N_T}}$), and just differ for the phase. 
The matrices $\mathbf{Q}_{ \rm RF}$ and $\mathbf{Q}_{\rm BB}$ can be found by using the Frobenius norm as a distance metric and solving the following optimization problem:
\begin{equation}
\begin{array}{llll}
 (\mathbf{Q}_{\rm RF}^*,\mathbf{Q}_{\rm BB}^*)= & \underset{\mathbf{Q}_{ \rm RF},\mathbf{Q}_{\rm BB}}{\arg\min} ||\mathbf{Q}^{\rm opt}-\mathbf{Q}_{\rm RF}\mathbf{Q}_{\rm BB}||_F \\ 
\text{subject to}  &
 |\mathbf{Q}_{\rm RF}(i,j)|=\frac{1}{\sqrt{N_T}}\; , \quad \forall i, j \\ &
  ||\mathbf{Q}_{\rm RF}\mathbf{Q}_{\rm BB}||_F^2 \leq M \;.
\end{array}
\label{eq:precoding_design}
\end{equation}
Similarly, with regard to the design of the post-coding beamforming matrix, the optimal FD beamformer $\mathbf{D}^{\rm opt}$ that we would use in case of no hardware complexity constraints is approximated by the product $\mathbf{D}_{\rm RF}\mathbf{D}_{\rm BB}$, where $\mathbf{D}_{\rm RF}$ is the $(N_R \times N_R^{RF})-$dimensional RF post-coding matrix and $\mathbf{D}_{\rm BB}$ is the $(N_R^{RF} \times M)-$dimensional baseband post-coding matrix. The entries of the RF post-coder $\mathbf{D}_{\rm RF}$  are constrained to have norm equal to $\frac{1}{\sqrt{N_R}}$. The matrices $\mathbf{D}_{\rm RF}$ and $\mathbf{D}_{\rm BB}$ can be then found solving the following optimization problem
\begin{equation}
\begin{array}{lll}
 (\mathbf{D}_{\rm RF}^*,\mathbf{D}_{\rm BB}^*)= & \underset{\mathbf{D}_{\rm RF},\mathbf{D}_{\rm BB}}{\arg\min} ||\mathbf{D}^{\rm opt}-\mathbf{D}_{\rm RF}\mathbf{D}_{\rm BB}||_F \\ 
\text{subject to} & |\mathbf{D}_{\rm RF}(i,j)|=\frac{1}{\sqrt{N_R}}\; , \quad \forall i, j  \; .
\end{array}
\label{eq:postcoding_design}
\end{equation}
It is easy to show that optimization problems \eqref{eq:precoding_design} and \eqref{eq:postcoding_design} are not convex optimization problem; inspired by  \cite{ghauch2016subspace}, we thus resort to the Block Coordinate Descent for Subspace Decomposition (BCD-SD) algorithm, that basically is based on a sequential iterative update of the analog part and of the baseband part of the beamformers. 
The algorithm's recipe,  whose complexity is tied to the third power of the
number of RF chains,  is reported in Algorithm \ref{BCD-SDA}. 
It is worth noting that the topic of HY beamforming is currently a very active research area, and several alternatives to the proposed solution are available. A detailed study about the performance of other HY beamforming schemes is however out of the scope of this paper; however, for benchmarking purposes, we will also show performance results for the case of FD beamforming, so as to quantify the loss incurred by the HY structures. 

\begin{algorithm}[!t]

\caption{Block Coordinate Descent for Subspace Decomposition Algorithm for Hybrid Beamforming}

\begin{algorithmic}[1]

\label{BCD-SDA}

\STATE Initialize $I_{\max}$ and set  $i=0$

\STATE Set arbitrary $\mathbf{Q}_{\rm{RF},0}$ and $\mathbf{D}_{\rm{RF},0}$

\REPEAT

\STATE  Update $\mathbf{Q}_{\rm{BB},i+1}=\left(\mathbf{Q}_{\rm{RF},i}^H\mathbf{Q}_{\rm{RF},i}\right)^{-1}\mathbf{Q}_{\rm{RF},i}^H\mathbf{Q}^{\rm opt}$ \\ and $\mathbf{D}_{\rm{BB},i+1}=\left(\mathbf{D}_{\rm{RF},i}^H\mathbf{D}_{\rm{RF},i}\right)^{-1}\mathbf{D}_{\rm{RF},i}^H\mathbf{D}^{\rm opt}$

\STATE Set $\phi_{i}=\mathbf{Q}^{\rm opt}\mathbf{Q}_{\rm{BB},i+1}^H\left(\mathbf{Q}_{\rm{BB},i+1}\mathbf{Q}_{\rm{BB},i+1}^H\right)^{-1}$ \\and $\psi_{i}=\mathbf{D}^{\rm opt}\mathbf{D}_{\rm{BB},i+1}^H\left(\mathbf{D}_{\rm{BB},i+1}\mathbf{D}_{\rm{BB},i+1}^H\right)^{-1}$

\STATE  Update $\mathbf{Q}_{\rm{RF},i}=\frac{1}{\sqrt{N_T}}e^{j\phi_{i}}$ \\and $\mathbf{D}_{\rm{RF},i}=\frac{1}{\sqrt{N_R}}e^{j\psi_{i}}$

\STATE Set  $i=i+1$

\UNTIL{convergence or $i=I_{\max}$}

\end{algorithmic}

\end{algorithm}

\subsection{HY beamforming for the MIMO-OFDM transceiver}
We now consider the issue of beamformer design for the MIMO-OFDM transceiver. From \eqref{eq:fftdomain_OFDM22} it is seen that the optimal pre-coders and post-coders for the detection of the data vector $\overline{\mathbf{S}}(n)$ are given by the left and right singular vectors associated to the $M$ largest eigenvalues of the matrix $ \overline{{\cal H}}(n)$, respectively. We will denote these optimal beamformers as $\mathbf{Q}^{\rm opt}(n)$ and $\mathbf{D}^{\rm opt}(n)$, respectively; differently from what happens for the SCM transceivers, these beamformers are now carrier dependent. Our aim is to approximate the optimal pre-coder  $\mathbf{Q}^{\rm opt}(n)$ with  the product $\mathbf{Q}_{ \rm RF}\mathbf{Q}_{\rm BB}(n)$, and the optimal post-coder $\mathbf{D}^{\rm opt}(n)$ with the product 
$\mathbf{D}_{ \rm RF}\mathbf{D}_{\rm BB}(n)$. 
Now, letting \cite{castanedaWSA2016}
$$
\begin{array}{lllll}
\mathbf{Q}^{\rm opt}= &\left[\mathbf{Q}^{\rm opt}(1), \ldots , \mathbf{Q}^{\rm opt}(k)\right] \; \; \in \mathbb{C}^{N_T \times kM}  \; ,
\\
\mathbf{D}^{\rm opt}= & \left[\mathbf{D}^{\rm opt}(1), \ldots , \mathbf{D}^{\rm opt}(k)\right] \; \; \in \mathbb{C}^{N_R \times kM} \; ,
\\
\mathbf{Q}_{\rm{BB}}= &\left[\mathbf{Q}_{\rm BB}(1), \ldots, \mathbf{Q}_{\rm BB}(k)\right] \; \; \in \mathbb{C}^{N_T^{RF} \times kM} \; ,
\\
\mathbf{D}_{\rm{BB}}= & \left[\mathbf{D}_{\rm BB}(1), \ldots, \mathbf{D}_{\rm BB}(k)\right] \; \; \in \mathbb{C}^{N_R^{RF} \times kM} \;, 
\end{array}
$$
the HY beamformer design amount to solving the following two constrained optimization problems
\begin{equation}
\begin{array}{lll}
 (\mathbf{Q}_{\rm{RF}}^*,\mathbf{Q}_{\rm {BB}}^*)= & \underset{\mathbf{Q}_{ \rm{RF}},\mathbf{Q}_{\rm{BB}}}{\arg\min} ||\mathbf{Q}^{\rm opt}-\mathbf{Q}_{\rm{RF}}\mathbf{Q}_{\rm{BB}}||_F \\
\text{subject to} & |\mathbf{Q}_{\rm{RF}}(i,j)|=\frac{1}{\sqrt{N_T}}  \;,  \\
& \|\mathbf{Q}_{\rm{RF}}\mathbf{Q}_{\rm{BB},k}\|_F^2 \leq kM \;,
\end{array}
\label{eq:precoding_design_multicarrier}
\end{equation}
and
 \begin{equation}
\begin{array}{llll}
 (\mathbf{D}_{\rm{RF}}^*,\mathbf{D}_{\rm{BB}}^*)= & \underset{\mathbf{D}_{\rm{RF}},\mathbf{D}_{\rm{BB}}}{\arg\min} ||\mathbf{D}^{\rm opt}-\mathbf{D}_{\rm{RF}}\mathbf{D}_{\rm{BB}}||_F \\
\text{subject to} & |\mathbf{D}_{\rm{RF}}(i,j)=\frac{1}{\sqrt{N_R}}\; .
\end{array}
\label{eq:postcoding_design_multicarrier}
 \end{equation}
The above optimization problems  have the same structure as problems in \eqref{eq:precoding_design} and \eqref{eq:postcoding_design}, and can thus be solved through a straightforward application of the BCD-SD algorithm. We do not explicitly report here the full details of the algorithm for the sake of brevity.

\section{Performance indicators}
The different transceiver architectures will be compared based on the ASE and the GEE.

\subsection{Computation of the ASE}
The ASE is the maximum achievable spectral efficiency with the constraint of arbitrarily small BER and of pre-fixed modulation type.  
The ASE takes the particular constellation and signaling parameters into consideration, so it does not qualify as a normalized capacity measure
(it is derived from the \textit{constrained capacity}). We focus here on ergodic rates so the {ASE} is computed given the channel realization  and averaged over it (remember that we are assuming perfect channel state information at the receiver). The spectral efficiency $\rho$ of any practical coded modulation system operating at a low packet error rate is upper bounded by the {ASE}, i.e., $\rho\leq \mathrm{ASE}$, where
\begin{equation}
\mathrm{ASE}=\frac{1}{T_{\rm s}W}\lim_{L\rightarrow\infty}\frac{1}{L} E_{\widetilde{\mathbf{H}}} \left[I(\mathbf{s};\hat{\mathbf{s}}|\widetilde{\mathbf{H}})\right]\;\mathrm{bit/s/Hz} \label{eq:ASE}
\end{equation}
$I(\mathbf{s};\hat{\mathbf{s}}|\tilde{\mathbf{H}})$ being the mutual information (given the channel realization) between the transmitted data symbols and their soft estimates,   $T_{\rm s}$ the symbol interval, and $W$ the signal bandwidth (as specified in Section~\ref{sec:numerical_results}). Although not explicitly reported, for notational simplicity, the ASE in (\ref{eq:ASE}) depends on the 
Signal-to-Interference plus Noise Ratio (SINR).

The computation of the mutual information requires the knowledge of the channel conditional probability density function (pdf) $p(\hat{\mathbf{s}}| \mathbf{s},\tilde{\mathbf{H}})$. As already said, it can be numerically computed by adopting the simulation-based technique described in~\cite{ArLoVoKaZe06} once the channel at hand is finite-memory and the optimal detector for it is available.
In addition, only the optimal detector for the actual channel is able to achieve the ASE in (\ref{eq:ASE}). 

In both transceiver models described in Section III the soft symbol estimates can be expressed in the form
\begin{equation}
\widehat{\mathbf{s}}(n)=\mathbf{C}\mathbf{s}(n)+ \sum_{\ell \neq 0} \mathbf{C}_{\ell} \mathbf{s}(n-\ell) + \mathbf{z}(n)
\label{eq:auxiliary}
\end{equation}
i.e., as a linear transformation (through matrix $\mathbf{C}$) of the desired QAM data symbols, plus a linear combination of the interfering data symbols and the colored noise $\mathbf{z}(n)$ having a proper covariance matrix. The optimal receiver has a computational complexity which is out of reach and for this reason we consider much simpler linear suboptimal receivers. Hence,  we are interested in the achievable performance when using suboptimal low-complexity detectors. We thus resort to the framework described in~\cite[Section VI]{ArLoVoKaZe06}. We compute proper lower bounds on the mutual information
 (and thus on the ASE) obtained by substituting $p(\hat{\mathbf{s}}| \mathbf{s},\tilde{\mathbf{H}})$ in the mutual information definition with an arbitrary auxiliary channel law $q(\hat{\mathbf{s}}| \mathbf{s},\tilde{\mathbf{H}})$
with the same input and output alphabets as the original channel (mismatched detection~\cite{ArLoVoKaZe06})---the more {accurately} the auxiliary channel {approximates} the actual one, the closer the bound is.
If the auxiliary channel law can be represented/described as a finite-state channel, the
pdfs $q(\hat{\mathbf{s}}|\mathbf{s}, \tilde{\mathbf{H}})$ and $q_p(\hat{\mathbf{s}}| \tilde{\mathbf{H}})=\sum_{\mathbf{s}}q(\hat{\mathbf{s}}|\mathbf{s}, \tilde{\mathbf{H}})P(\mathbf{s})$
can be computed, this time, by using the optimal maximum a posteriori symbol detector
for that auxiliary channel \cite{ArLoVoKaZe06}. This detector, that
is clearly suboptimal for the actual channel, has at its
input the sequence $\hat{\mathbf{s}}$ generated by simulation \emph{according
to the actual channel model} (for details, see \cite{ArLoVoKaZe06}).  If we
change the adopted receiver (or, equivalently, if we change the auxiliary
channel) we obtain different lower bounds on the constrained capacity
but, in any case, these bounds are \emph{achievable} by those receivers,
according to mismatched detection theory~\cite{ArLoVoKaZe06}.
We thus say, with a slight abuse of terminology, that the computed
lower bounds are the ASE values of the considered channel when those
receivers are employed. 
This technique thus allows us to take reduced-complexity receivers into account. In fact, it is sufficient to consider an auxiliary channel which is a simplified version of the actual channel in the sense that
only a portion of the actual channel memory and/or a limited number of impairments are present. In particular, we will use the auxiliary channel law (\ref{eq:auxiliary}), where the sum of the interference and the thermal noise
$\mathbf{z}(n)$ is assimilated to Gaussian noise with a proper covariance matrix.

The transceiver models are compared in terms of ASE without taking into account specific coding schemes, being understood that, with a properly designed channel code, the information-theoretic performance can be closely approached. 

\subsection{GEE and power consumption models}
The GEE is defined as the ratio between the achievable rate and the overall power consumption, taking into account both the
radiated power and the  power consumed by the hardware circuitry\cite{buzzi2016survey}. GEE is  measured in [bit/Joule] and is expressed as: 
\begin{equation}
{\rm GEE}= \displaystyle \frac{ W {\rm ASE}}{\beta P_T + P_{\rm{TX},c}+ P_{\rm{RX},c}} \; ,
\label{eq:GEE}
\end{equation}
where  $W$ is the system bandwidth,  $P_{\rm{TX},c}$ is the amount of power consumed by the transmitter circuitry,  $P_{\rm{RX},c}$ is the amount of power consumed by the receiver's device circuitry, and 
$\beta>1$ is a scalar coefficient modeling the power amplifier inefficiency. Note that, differently from what happens in the most part of existing studies on energy efficiency for cellular communications, we have included in the GEE definition \eqref{eq:GEE}  the power consumed both at the transmitter and at the receiver's devices.

\subsubsection{Transceiver power consumption for FD beamforming}
In the case of SCM-TDE the considered FD pre-coding architecture requires a baseband digital precoder that adapts the $M$ data streams  to the $N_T$ transmit antennas; then, for each antenna there is a  digital-to-analog-converter (DAC), an RF chain and a power amplifier (PA). At the receiver, for the SCM-TDE,  a low noise amplifier (LNA), an RF chain, an analog-to-digital converter (ADC) is required for each antenna, plus a baseband digital combiner that combines the $N_R$ outputs of ADC to obtain the soft estimate of the $M$ trasmitted symbols. 
Following thus the same approach as in \cite{Buzzi_Beamforming2016}, 
the amount of power consumed by the transmitter circuitry can be expressed as
$P_{\rm{TX},c}=N_T\left(P_{\rm RFC}+P_{\rm DAC}+P_{\rm PA}\right)+P_{\rm BB}$,
and the amount of power consumed by the receiver circuitry can be expressed as
$P_{\rm{RX},c}=N_R\left(P_{\rm RFC}+P_{\rm ADC}+P_{\rm LNA}\right)+P_{\rm BB}$.
According to \cite{Buzzi_Beamforming2016} and references therein, $P_{\rm RFC}= 40$ mW  is the power consumed by the single RF chain, $P_{\rm DAC}= 110$ mW is the power consumed by each DAC, 
$P_{\rm ADC}=200$mW   is the power consumed by each single ADC, 
$P_{\rm PA}=16 $ mW  is the power consumed by the PA,
$P_{\rm LNA}=30$ mW  is the power consumed by the LNA,  
and $P_{\rm BB}$ is the amount of power consumed by the baseband precoder/combiner designed; assuming a  CMOS implementation we have a power consumption of 243 mW.

In the case of SCM-FDE the transmitter has the same power consumption of the previous case. At the receiver, for the SCM-FDE,  after the baseband digital combiner there are the operations of FFT and IFFT in order to implement the frequency domain equalization. The amount of power consumed by the transmitter circuitry can be thus expressed as
$
P_{\rm{TX},c}=N_T\left(P_{\rm RFC}+P_{\rm DAC}+P_{\rm PA}\right)+P_{\rm BB}$,
and the amount of power consumed by the receiver circuitry can be expressed as
$P_{\rm{RX},c}=N_R\left(\!P_{\rm RFC}\!+\!P_{\rm ADC}\!+\!P_{\rm LNA}\!+\! P_{c,\rm{FFT}}\!+\!P_{c,\rm{IFFT}}\right) +P_{\rm BB}$.
We have already provided numerical values for the above quantities, except that for $P_{c,\rm{FFT}}$ and for $P_{c,\rm{IFFT}}$ that are the powers consumed by the FFT and IFFT processing. For  a 256-point FFT and IFFT\footnote{In the numerical simulations we will be taking $k=256$.} we have $P_{c,\rm{FFT}}=153$ mW and $P_{c,\rm{IFFT}}=153$ mW \cite{kannan2009hardware}.

Similarly, for the OFDM transceiver we have
$P_{\rm{TX},c}=N_T\left(P_{\rm RFC}+P_{\rm DAC}+P_{\rm PA}+P_{c,\rm{IFFT}}\right)+P_{\rm BB}$, and
$P_{\rm{RX},c}=N_R\left(P_{\rm RFC}+P_{\rm ADC}+P_{\rm LNA}+ P_{c,\rm{FFT}}\right)+P_{\rm BB}$.

\subsubsection{Transceiver power consumption for HY beamforming}
For HY beamforming and SCM-TDE transceiver, the amount of power consumed by the transmitter circuitry is expressed as
$P_{\rm{TX},c}=N_T^{\rm RF}\left(P_{\rm RFC}+P_{\rm DAC}+N_T P_{\rm PS}+P_{\rm PA}\right)+P_{\rm BB}$, 
while the amount of power consumed by the receiver circuitry is 
$P_{\rm{RX},c}=N_R^{\rm RF}\left(P_{\rm RFC}+P_{\rm ADC}+N_R P_{\rm PS}\right)+N_R P_{\rm LNA}+P_{\rm BB}$.
Numerical values for the above quantities have already been provided, except that for $P_{\rm PS}$, the power consumed by each  phase shifters, that we assume to be 30 mW.
In the case of SCM-FDE, we have
$P_{\rm{TX},c}=N_T^{\rm RF}\left(P_{\rm RFC}+P_{\rm DAC}+N_T P_{\rm PS}+P_{\rm PA}\right) +P_{\rm BB}$,
and 
$P_{\rm{RX},c}=N_R^{\rm RF}\!\left(\!P_{\rm RFC}\!+\!P_{\rm ADC}\!+\!N_R P_{\rm PS}\!+\!P_{c,\rm{FFT}}  +\!P_{c,\rm{IFFT}}\!\right)+N_R P_{\rm LNA}+P_{\rm BB}$.
Finally, for the MIMO-OFDM case we have 
$P_{\rm{TX},c}=N_T^{\rm RF}\left(\!P_{\rm RFC}\!+\!P_{\rm DAC}\!+\!N_T P_{\rm PS}\!+\!P_{\rm PA}\!+\!P_{c,\rm{IFFT}}\!\right) \!+\!P_{\rm BB} $, and
$P_{\rm{RX},c}=N_R^{\rm RF}\left(P_{\rm RFC}+P_{\rm ADC}+N_R P_{\rm PS}+P_{c,\rm{FFT}}\right)+N_R P_{\rm LNA}+P_{\rm BB}$.

\begin{figure}[t]
\centering
\includegraphics[scale=0.26]{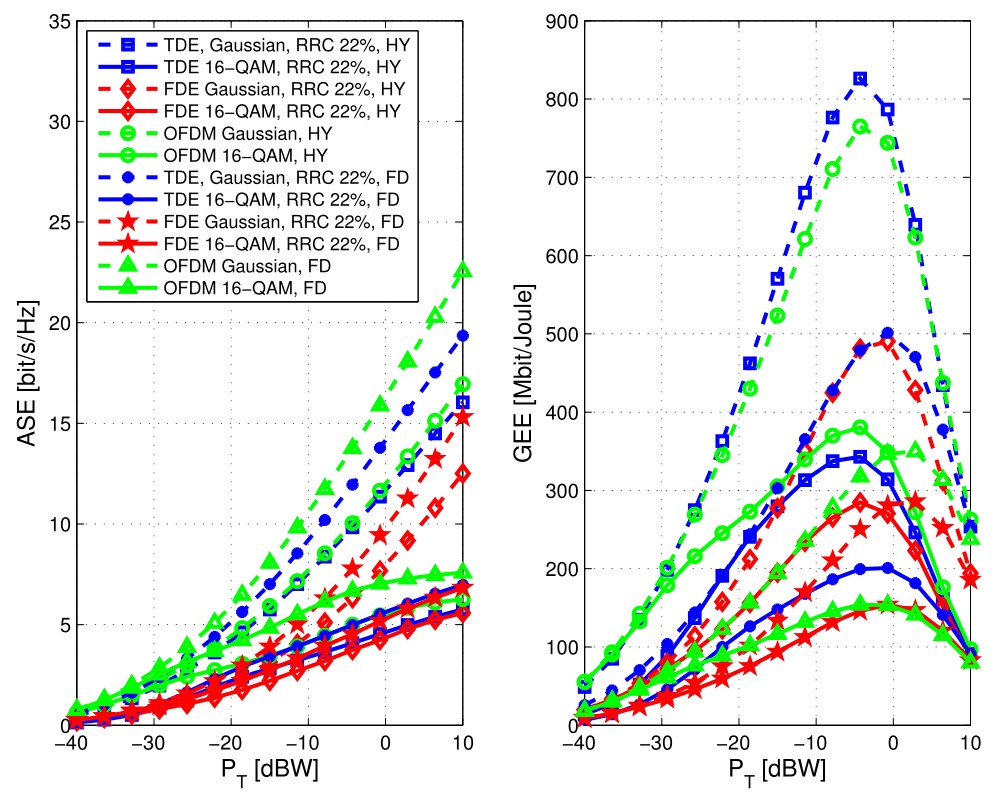}
\caption{ASE and GEE versus transmit power; comparison of TDE, FDE and OFDM, with finite and infinite modulation cardinality and with comparison of HY and FD beamforming and linear PA. Parameters: $M=2$; $d=30$ m $N_R \times N_T= 10 \times 50$, in the case of HY beamforming $N_T^{\rm RF}=N_R^{\rm RF}=2$.}
\label{Fig:ASE_GEE_linear}
\end{figure} 

\begin{figure}[t]
\centering
\includegraphics[scale=0.26]{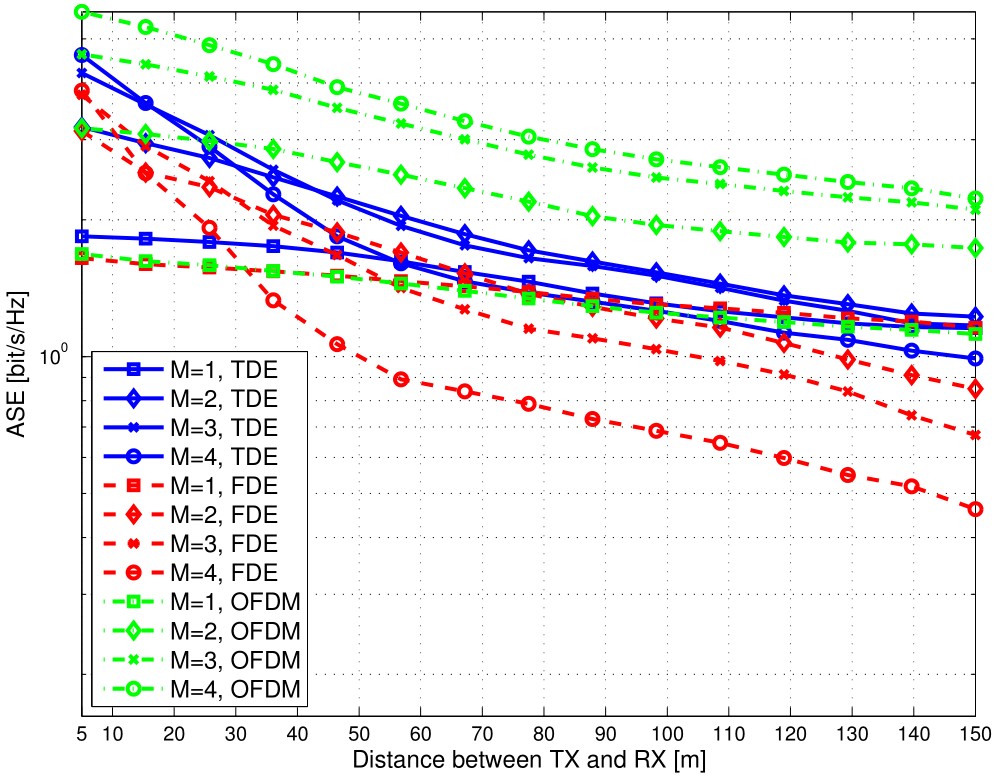}
\caption{ASE versus distance; impact of multiplexing order $M$ with TDE, FDE and OFDM and linear PA. Parameters: 4-QAM modulation; $P_T=0$ dBW; $N_R \times N_T= 10 \times 50$, HY beamforming with $N_T^{\rm RF}=N_R^{\rm RF}=M$.}
\label{Fig:ASE_distance_M_linear}
\end{figure}

\section{Numerical Results} \label{sec:numerical_results}
In our simulation setup, we consider a communication bandwidth of $W=500$ MHz centered over a mmWave carrier frequency. The MIMO propagation channel, described in Section II,  has been generated according to the statistical procedure detailed in \cite{buzzidandreachannel_model}. The additive thermal noise is assumed to have a power spectral density of -174 dBm/Hz, while the front-end receiver is assumed to have a noise figure of 3 dB. 
For SCMs, the Square Root Raised Cosine (SRRC) pulse with roll-off factor 0.22 is adopted for the shaping transmit and receive filters.
For this waveform, we define the bandwidth as the frequency range such that out-of-band emissions are 40 dB below the maximum in-band value of the Fourier transform of the pulse.  For the considered communication bandwidth of $W=500$ MHz, we found that the symbol interval $T_{\rm s}$ is 2.15 ns,  for the case in which we consider its truncated version to the interval $[-4T_{\rm s}, 4T_{\rm s}]$. 
For the OFDM case, we are instead considering rectangular pulses of duration $T_s=1/W$; in this case, we show in the figures the 90$\%$ value of the obtained ASE to take into account the fact that carriers at the edge of the frequency band are usually not loaded to reduce out-of-band emissions. Moreover, for the SCM-FDE and OFDM schemes the reported ASE values include the penalty factor $(1-C/k)$ due to the insertion of the cyclic prefix, with $k=256$. We also point out that
the reported results are to be considered as an ideal benchmark both for the ASE and the GEE since we are considering a single-user link and we are neglecting the interference.\footnote{We note however that being mmWave systems mainly noise-limited rather than interference limited, the impact of this assumption on the obtained results is very limited.} HY pre-coding and post-coding, with  $M$ RF chains at the transmitter and at the receiver, is considered, also in comparison to FD structures. All the figures refer to the case that $N_R \times N_T= 10 \times 50$, and uniform linear arrays were used. 

We start by considering the case that the power amplifier operates in the linear regime and causes no distortion.
Fig. \ref{Fig:ASE_GEE_linear} reports the ASE\footnote{Of course, the achievable rates in bit/s can be immediately obtained by multiplying the ASE by the communication bandwidth $W=500$ MHz.} and the GEE for  SCM-TDE, SCM-FDE and MIMO-OFDM transceivers versus the transmit power $P_T$. Both the cases of finite cardinality (16-QAM) data symbols and Gaussian data symbols are considered. Fig. \ref{Fig:ASE_distance_M_linear} reports the ASE for the three considered transceivers versus the link length $d$, assuming that the transmit power is $P_T=0$ dBW. While  Fig. \ref{Fig:ASE_GEE_linear} contains a comparison between the 16-QAM modulation scheme and the case of Gaussian-distributed data symbols, Fig.  \ref{Fig:ASE_distance_M_linear} focuses on the case of 4-QAM modulation and studies the impact of the multiplexing order $M$. 
 
We can see that there are no huge differences among the performance of the considered transceivers, even though the OFDM appears to be the best-performing scheme, while SCM-FDE 
achieves the worst performance.\footnote{However, as reported in \cite{buzzi2016spectral}, SCM-TDE turns out to be the best performing scheme for the case in which a linear minimum mean square error equalizer is used in turn of the ZF equalizer. Additionally, results not reported here due to lack of space have shown that the performance of SCM-TDE can be improved considerably by considering fractionally-spaced TDE.} While, as expected, FD beamforming outperforms HY beamforming in terms of ASE, we see that in terms of GEE the reverse is generally true. As an instance, if we focus on the OFDM transceiver with 16-QAM modulations and with a transmit power of 0 dBW, it is seen that the HY beamformer exhibits a loss in terms of ASE with respect to the FD beamformer on the order of about 22$\%$, while, instead, its gain, in terms of GEE, is around $120 \%$. The advantage of HY structures with respect to FD structures in terms of GEE can however disappear for MIMO links with large number of antennas, as reported in \cite{Buzzi_Beamforming2016}.

\begin{figure}[t]
\centering
\includegraphics[scale=0.26]{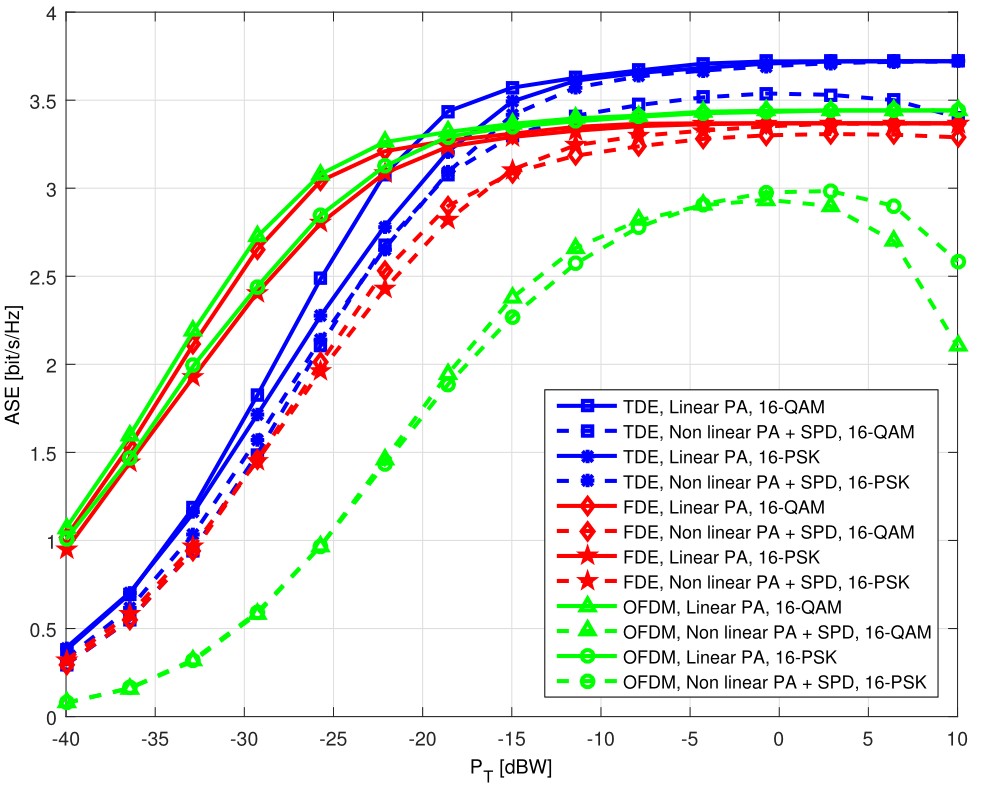}
\caption{ASE versus transmit power; comparison of TDE, FDE and OFDM, with 16-QAM and 16-PSK modulations, linear and non linear PA. Parameters: $M=1$; $d=30$ m $N_R \times N_T= 10 \times 50$, HY beamforming with $N_T^{\rm RF}=N_R^{\rm RF}=1$.}
\label{Fig:fig_16_QAM_PSK}
\end{figure}

\begin{figure}[t]
\centering
\includegraphics[scale=0.26]{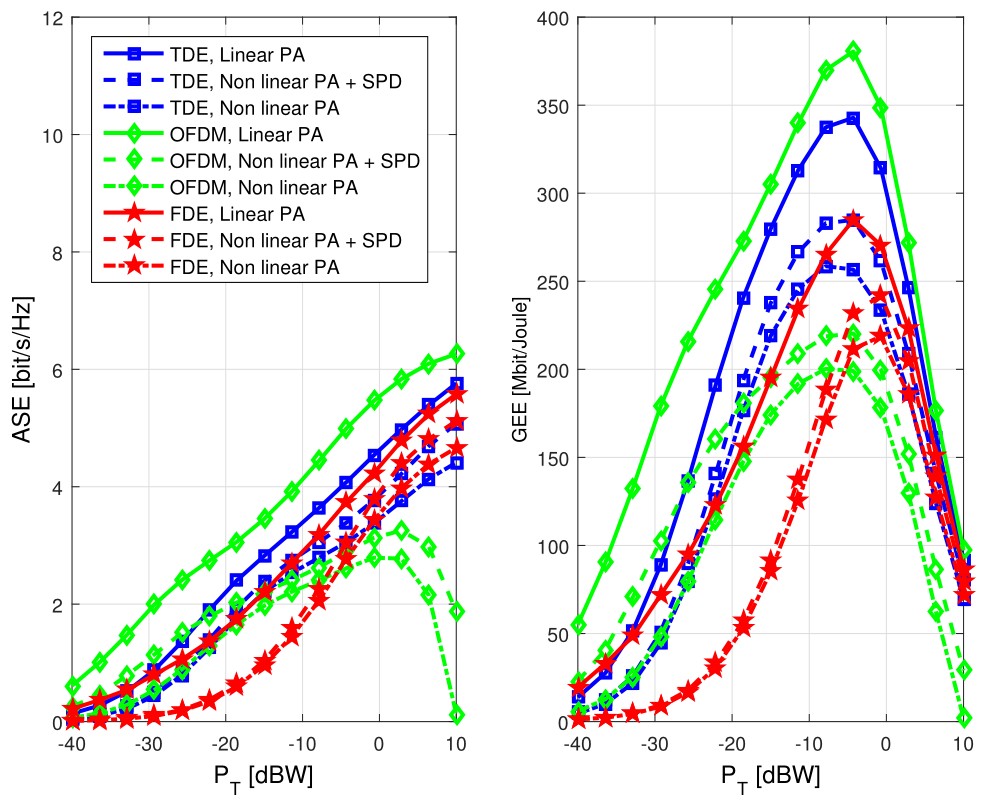}
\caption{ASE and GEE versus transmit power for the case in which non-linearities in the power amplifier are taken into account. Parameters: $d=30$ m; $M=2$; $N_R \times N_T= 10 \times 50$; HY beamforming with $N_T^{\rm RF}=N_R^{\rm RF}=2$, 16-QAM modulation.}
\label{Fig:fignonlinearPt}
\end{figure}

\begin{figure}[t]
\centering
\includegraphics[scale=0.26]{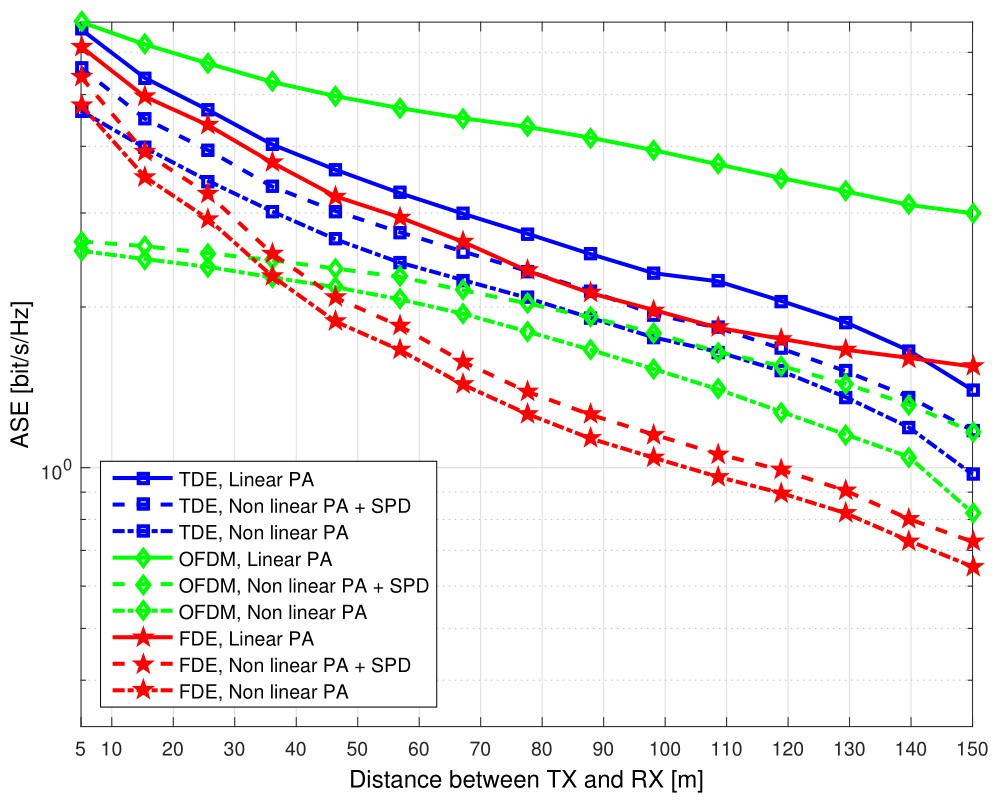}
\caption{ASE versus distance for the case in which non-linearities in the power amplifier are taken into account.
Parameters: $P_T=0$ dBW; $M=2$; $N_R \times N_T= 10 \times 50$; HY beamforming with $N_T^{\rm RF}=N_R^{\rm RF}=2$, 16-QAM modulation.}
\label{Fig:fignonlineard}
\end{figure}

In order to have an insight into the effect of power amplifiers non-linearities, Figs. \ref{Fig:fig_16_QAM_PSK}, \ref{Fig:fignonlinearPt} and \ref{Fig:fignonlineard}  show the system ASE, when the power amplifier non-linearities are taken into account.  Fig.  \ref{Fig:fig_16_QAM_PSK} also reports a comparison between 16-QAM modulation and the constant-envelope 16-PSK modulation, while Fig. \ref{Fig:fignonlinearPt} also shows GEE results.
It is seen from the figure that the non-linear behavior of the power amplifier introduces some performance degradation, especially on the OFDM scheme, which now performs worse than the other two schemes, while instead SCM-TDE reveal to be the best performing scheme.  The results thus confirm that also for mmWave frequencies the OFDM scheme is more sensitive than SCM schemes to the peak-to-average power ratio (PAPR) problem. 
Moreover, the fact that  SCM-FDE performs worse than OFDM can be intuitively justified by noting  that OFDM uses a subcarrier-dependent digital pre-coder and post-coder, which is not the case for SCM-FDE.

\begin{figure}[t]
\centering
\includegraphics[scale=0.26]{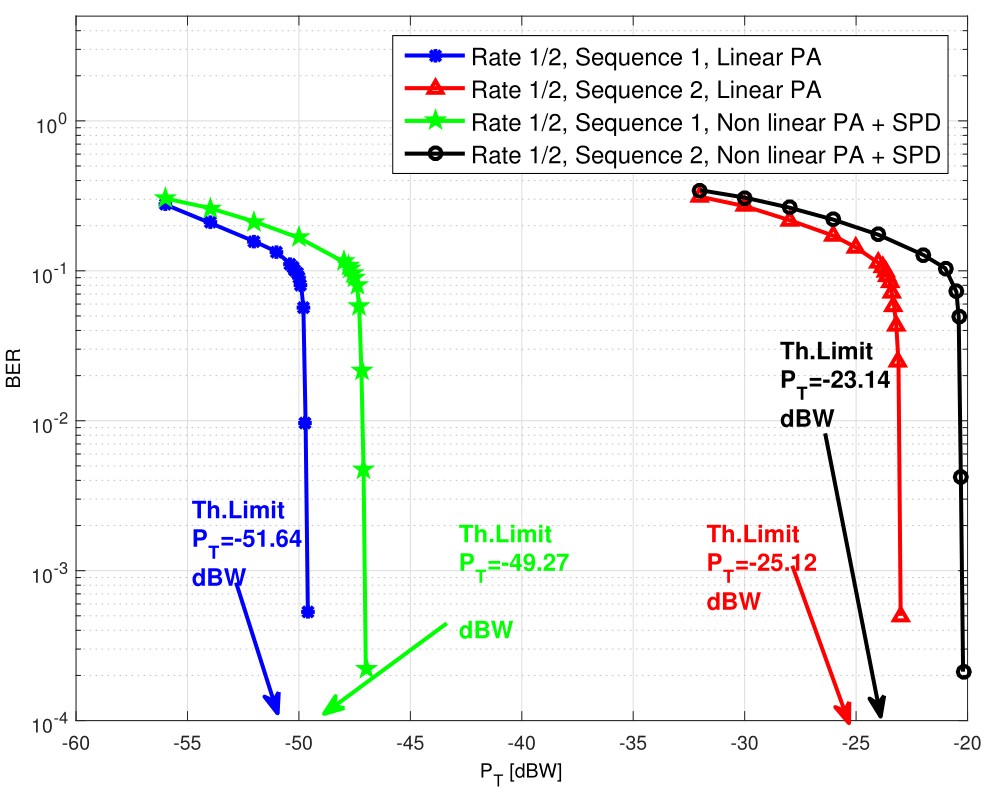}
\caption{BER of TDE for 16-QAM, d=30 m, $M$=2, $N_R \times N_T= 10 \times 50$, HY beamforming with $N_T^{\rm RF}=N_R^{\rm RF}=2$.}
\label{Fig:fig_BER_TDE}
\end{figure}

\begin{figure}[t]
\centering
\includegraphics[scale=0.26]{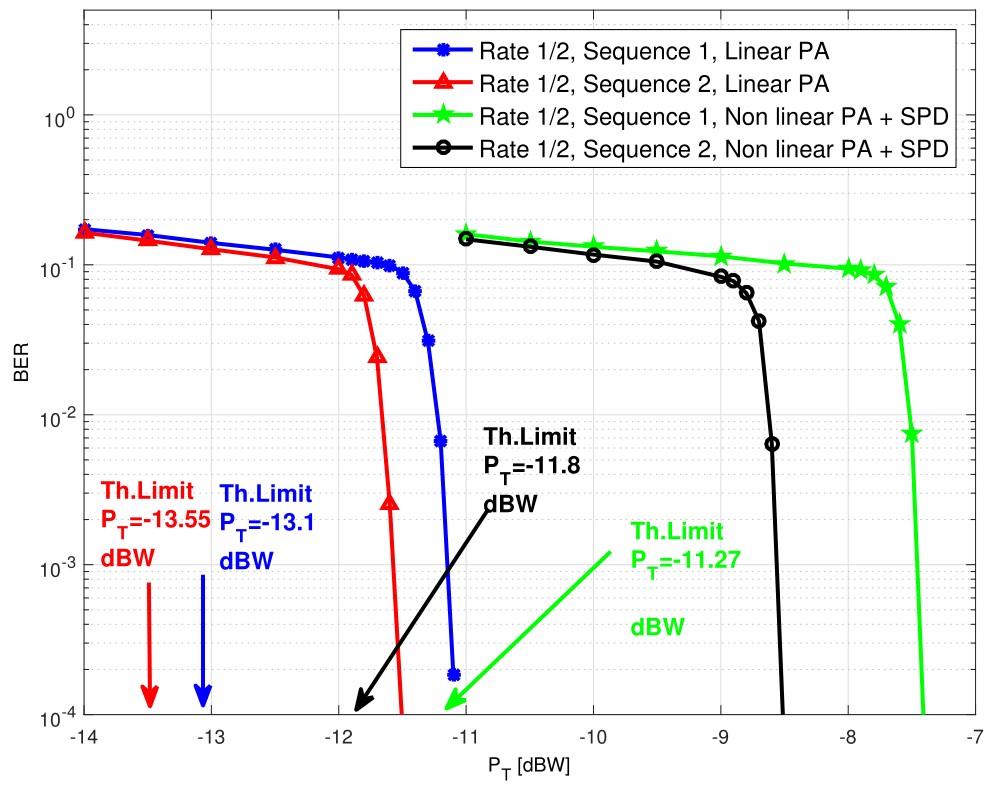}
\caption{BER of FDE for 16-QAM, d=30 m, $M$=2, $N_R \times N_T= 10 \times 50$, HY beamforming with $N_T^{\rm RF}=N_R^{\rm RF}=2$.}
\label{Fig:fig_BER_FDE}
\end{figure}

\begin{figure}[t]
\centering
\includegraphics[scale=0.26]{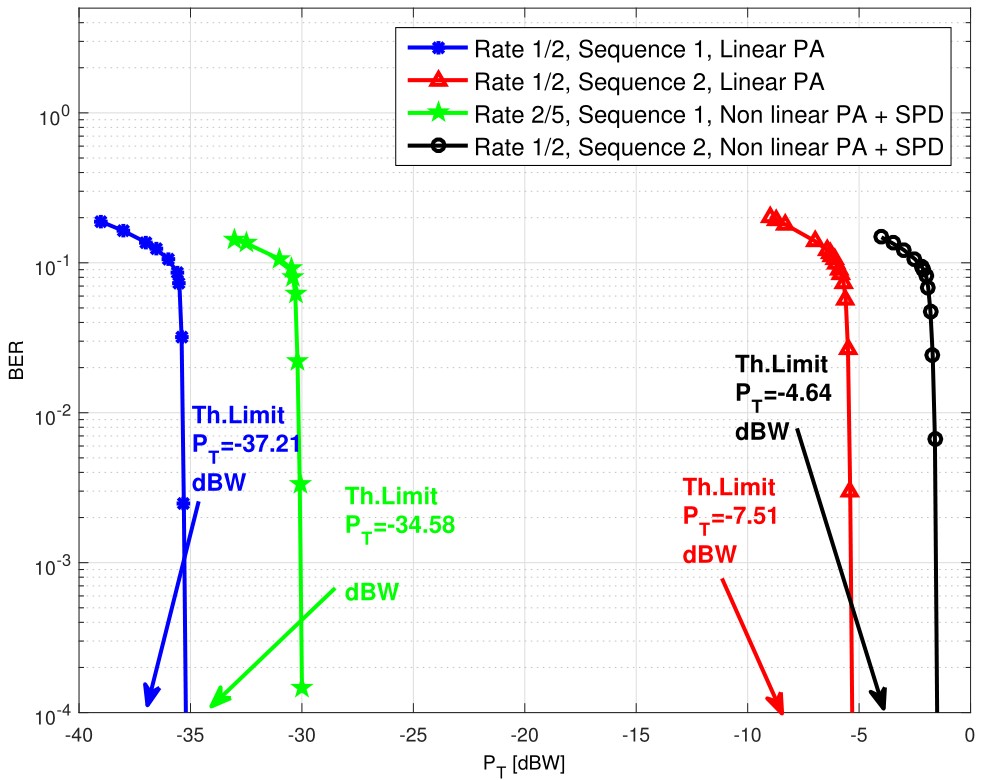}
\caption{BER of OFDM for 16-QAM, d=30 m, $M$=2, $N_R \times N_T= 10 \times 50$, HY beamforming with $N_T^{\rm RF}=N_R^{\rm RF}=2$.}
\label{Fig:fig_BER_OFDM}
\end{figure}

Other general comments about the obtained results are in the following.
\begin{itemize}
\item[-]
Results, in general,  improve for increasing transmit power and  for decreasing distance $d$ between transmitter and receiver.
\item[-]
In particular, a good performance can be attained for distances up to 100 m, whereas for $d>100$ m we have 
a steep degradation of the ASE. In this region, all the advantages given by increasing the modulation cardinality or the multipexing gain $M$ are essentially lost or reduced at very small values. Of course, this performance degradation may be compensated by increasing the transmit power:this thus confirms that mmWave are well suited for wireless cellular communications for distances up to 100 m.
\item[-]
For a reference distance of 30 m (which will be a typical one in small-cell 5G mmWave deployments for densely crowded areas), a {transmit} power around 0 dBW is enough to grant good performance and to benefit from the advantages of increased modulation cardinality, size of the antenna array, and multiplexing order.
\end{itemize}
We now proceed to showing BER results.
In Figs~\ref{Fig:fig_BER_TDE}, \ref{Fig:fig_BER_FDE} and \ref{Fig:fig_BER_OFDM} we report the BER results respectively of 16-QAM SCM-TDE, SCM-FDE and MIMO-OFDM when employing low-density-parity-check (LDPC) codes of rate equal to 1/2 and 2/5, in order to show how practical (i.e., finite-length and not \textit{ad hoc} designed) codes perform in one realization of the considered scenario, which entails {$M=2$, $d=30$ m, $N_R \times N_T=10 \times 50$}. The parameters of the codes are reported in Table \ref{tab:code_par} where $r_c$ denotes the rate of the code and the degree distributions of variable and check nodes are provided by giving the fraction $a_{i}$ ($\sum_{i}a_{i}=1$) of degree $i$ nodes. In any case, the codeword length is $L=64800$ bits, and the decoder iterations are limited to 40. These codes were designed for low intersymbol interference (ISI) channels, and, despite not specifically designed for these systems, they closely approach the provided ASE lower bounds. Since with {$M=2$} the two multiplexed streams perform differently, the code rates on each stream should be tailored accordingly.

\begin{table*}
\protect\caption{Code rates and degree distributions of the employed LDPC codes.\label{tab:code_par}}

\centering{}%
\begin{tabular}{c c c}
\hline 
\hline
$r_c$ & variable node distribution & check node distribution\tabularnewline
\hline 
1/2 & {\scriptsize{}$\begin{array}{ccc}
a_{2}=0.499985 & a_{3}=0.3 & a_{8}=0.200015\end{array}$} & {\scriptsize{}$\begin{array}{cc}
a_{7}=0.999815 & a_{8}=0.000185185\end{array}$}\tabularnewline

2/5 & {\scriptsize{}$\begin{array}{ccc}
a_{2}=0.599985 & a_{3}=0.267006 & a_{12}=0.133009\end{array}$} & {\scriptsize{}$\begin{array}{cc}
a_{5}=0.00483539 & a_{6}=0.995165\end{array}$}\tabularnewline
\hline
\hline 
\end{tabular}
\end{table*}

\section{Conclusion}
This paper has provided a comparison between single-carrier modulation schemes and conventional OFDM for a MIMO link operating at mmWave frequencies. In particular, two SCM techniques have been considered, SCM-TDE and SCM-FDE, and these transceivers have been compared with the MIMO-OFDM scheme. Our analysis has jointly taken into account  the modulation finite cardinality, the peculiarity of the channel matrix at mmWave frequencies (a clustered model has been adopted), the adoption of HY analog/digital beamforming structures, and the effect of hardware non-idealities such as the non-linear behavior of the transmit power amplifiers. 
Results have shown that the SCM-TDE structure achieves superior performance with respect to the other two competing schemes, in particular when non-linear distortions are taken into account. The advantage of HY beamforming structures with respect to FD ones in terms of GEE has also been highlighted. 

The present study can be generalized and strengthened  in many directions. In particular, the considered analysis might be applied in a multiuser environment; additionally, since, as already discussed, the reduced wavelength of mmWave frequencies permits installing arrays with many antennas in small volumes, an analysis, possibly through asymptotic analytic considerations, of the very large number of antennas regime could also be made. Finally, the impact of  the use of non-linear equalization schemes might also be investigated.

\bibliographystyle{IEEEtran}

\bibliography{finalRefs}

\end{document}